\documentclass[reprint,aps,pre]{revtex4-2}

\usepackage{amsmath,amssymb}
\usepackage{bm}

\usepackage{graphicx}
\usepackage{multirow}
\usepackage{dcolumn}
\usepackage{float}

\usepackage[dvipsnames]{xcolor}

\usepackage{hyperref}
\hypersetup{
    colorlinks=true,
    citecolor=red,
    linkcolor=NavyBlue,
    urlcolor=NavyBlue
}

\begin{document}
\title{Chimera States in Wheel Networks}
\author{Medha Bhindwar${}^{3,2}$}
\thanks{These authors contributed equally to this work.}

\author{Ashwathi Poolamanna${}^{2,4,5}$}
\thanks{These authors contributed equally to this work.}

\author{Yuvrajsingh Pravinsingh Patil${}^{1}$}
\author{Chandrakala Meena${}^{1,2}$}
\thanks{Contact author:chandrakala@iiserpune.ac.in}
\affiliation{${}^{1}$ Indian Institute of Science Education and Research (IISER), Pune, 411008, India}

\affiliation{${}^{2}$ Indian Institute of Science Education and Research (IISER), Thiruvananthapuram, 695551, India}
\affiliation{${}^{3}$ National Brain Research Centre, Gurugram, India, 122052}
\affiliation{${}^{4}$ Heyrovsky Institute, Czech Academy of Sciences, Dolejskova 2155/3, 182 00 Prague 8, Czech Republic}
\affiliation{${}^{5}$ Faculty of Mathematics and Physics, Charles University, Ke Karlovu 3, Praha, 12116, Czech Republic}
\date{\today}

\begin{abstract}
     Chimera states, characterized by the coexistence of coherent and incoherent dynamics, are a hallmark of symmetry-breaking phenomena in networks of coupled oscillators. Here, we investigate the role of higher-order interactions in the emergence of chimera states in wheel networks of R\"{o}ssler and Lorenz oscillators with pairwise and higher-order interactions under diffusive, conjugate, and mean-field diffusive coupling. Using multiple random initial conditions, we quantify the prevalence of synchronized, desynchronized, chimera, and synchronized cluster states. Chimera states emerge across all three coupling mechanisms as symmetry-broken states among the structurally and dynamically equivalent peripheral oscillators. Higher-order interactions substantially reorganize the dynamical landscape; in particular, they shift chimera states toward weaker pairwise coupling under diffusive coupling, predominantly support them at moderate coupling strengths under conjugate coupling, and yield comparatively less prevalent chimera states under mean-field diffusive coupling. These effects depend strongly on the intrinsic oscillator dynamics, particularly at larger higher-order coupling strengths. Overall, higher-order interactions neither universally promote nor suppress chimera states; rather, their influence is governed by the interplay between network symmetry, coupling mechanism, and intrinsic oscillator dynamics. The observed collective states are quantitatively characterized using two proposed statistical measures, while master-stability-function analysis provides a stability-based explanation for the loss of complete synchronization and the emergence of nonsynchronous dynamical regimes.

\end{abstract}

\maketitle

\section{INTRODUCTION}
\label{sec:1}
\noindent 
Understanding the dynamics of complex systems is fundamental to uncovering the mechanisms underlying diverse natural and engineered networks, including social networks\cite{cencetti2021temporal,rai2025ipsr}, structural and functional brain networks \cite{lv2021functional, pathak2022whole, wang2024multi}, other biological networks \cite{mollison1977spatial, kiss2017mathematics, gomez2023new}, and climate networks \cite{meena2017effect}. These systems are often modeled as dynamical networks, where nodes represent interacting elements and edges denote pairwise interactions\cite{boccaletti2006complex, boccaletti2023structure}. While this framework has been fundamental in advancing our understanding of complex systems, it has become increasingly evident that many real-world networks, ranging from ecological and social to neuronal systems, exhibit higher-order interactions (HOIs) where more than two units interact simultaneously \cite{ petri2014homological, singh2024higher, gambuzza2021stability, boccaletti2023structure, singh2026chimera}. Such interactions are modeled through hypergraphs and simplicial complexes, which extend traditional pairwise models to capture group dynamics in a systematic manner.

Coupled oscillator networks provide a prototypical setting for uncovering emergent collective behavior, such as synchronization, desynchronization, synchronization clusters, and chimera states \cite{starnetwork,wang2025stability}. Chimera states are a symmetry-breaking phenomenon where synchronized and desynchronized dynamics coexist within the same system \cite{mishra2023chimeras, kuramoto, abrams, zhu2014chimera, kemeth2016classification, parastesh2021chimeras, kundu2022higher,wang2024multi}. Although chimeras have been extensively studied in pairwise-coupled networks \cite{ulonska2016chimera, panaggio2016chimera, majhi2019chimera}, their manifestation under HOIs has only recently attracted attention, with investigations in nonlocal \cite{ghosh2024chimeric, wang2024multi} and globally coupled frameworks \cite{li2023chimera, kundu2022higher}.

Recent studies demonstrate that higher-order interactions promote the emergence of chimera states in networks of coupled phase oscillators \cite{kundu2022higher, kar2024effect, jaros2023higher,li2025analysis}. Similar effects are reported for other oscillator systems, including Stuart–Landau oscillators \cite{muolo2024phase, singh2024higher, singh2026chimera}, swarmalators \cite{hu2025effect} prey–predator Rosenzweig–MacArthur models \cite{ghosh2024chimeric}, and neuronal dynamics \cite{majhi2025patterns, wang2024multi,hu2024role}. Despite these advances, the role of higher-order interactions in shaping chimera states in networks of chaotic oscillators remains largely unexplored. Chaotic dynamical systems exhibit a rich spectrum of behaviors, ranging from fixed points and periodic orbits to chaos \cite{thakur2024machine,bhindwar2024role}, and their collective dynamics differ substantially from those of phase- or weakly nonlinear oscillators. Consequently, investigating how higher-order interactions influence the emergence and robustness of chimera states in networks of coupled chaotic oscillators constitutes a natural and important extension of existing studies.
 
Motivated by this perspective, in this study, we examine chimera states in a wheel network of coupled chaotic oscillators. The wheel network structure is closely related to the star network. Star networks, widely studied as canonical motifs in computer and scale-free networks, show various diverse dynamical behaviors, including chimeras, under pairwise diffusive, conjugate, and mean-field diffusive coupling schemes \cite{starnetwork}. The wheel topology extends this framework by incorporating nearest-neighbor interactions among peripheral nodes in addition to central-peripheral connections. This additional connectivity naturally introduces three-way interactions, which can be formally interpreted as 2-simplices within simplicial complexes \cite{zhang2023higher, li2023chimera}.
Many studies suggest that collective dynamics of dynamical networks depend on network structure and interaction mechanisms \cite{meena2020resilience, skardal2020higher, zhang2022higher, meena2023emergent}. Therefore, in this work, we consider the dynamics of R\"{o}ssler and Lorenz oscillators on the nodes of a wheel network incorporating higher-order interactions in addition to pairwise connections. We consider three coupling mechanisms: diffusive, conjugate, and mean-field diffusive coupling. Thus, we obtain six different dynamical configurations and systematically investigate the  collective dynamical behavior that emerge at the peripheral nodes of the wheel network, by varying the pairwise and higher-order interaction strengths. To distinguish among synchronized, desynchronized, chimera, and cluster states, we introduce suitable statistical measures for classifying the different dynamical behaviors. Furthermore, to assess the robustness of these states, we perform extensive simulations over a range of random initial conditions and evaluate the corresponding probabilities associated with each dynamical regime, with particular emphasis on chimera states. 
%These contrasting outcomes highlight the importance of exploring dynamical behaviors in dynamical networks which incorporates HOIs.

\begin{figure}[htbp]
    \centering
    \includegraphics[width=1\linewidth]{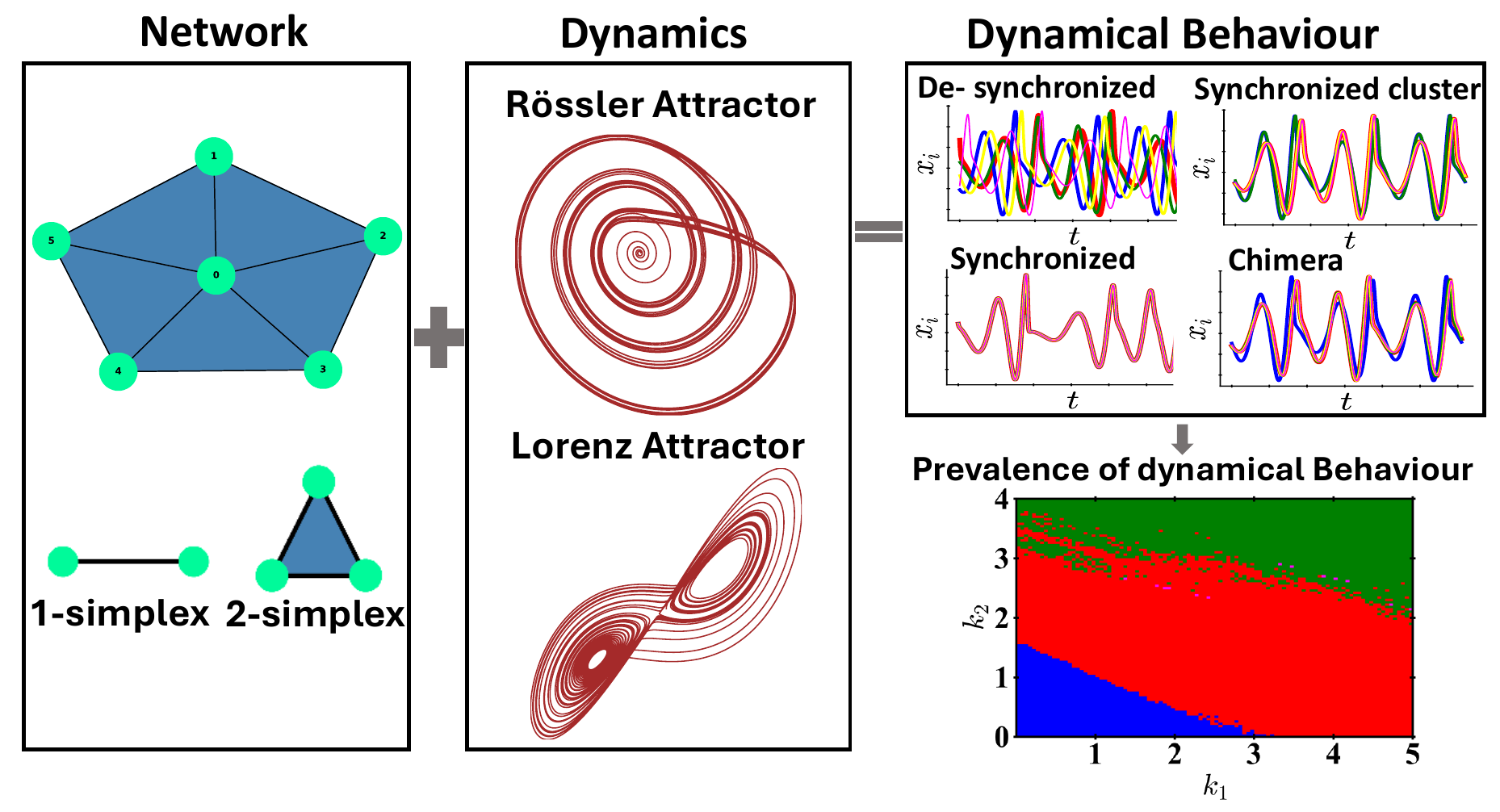}
    %\vspace{-6cm}
    \caption{\textbf{Schematic illustration of the workflow used to identify dynamical behaviors and analyze their robustness.} A wheel network with both pairwise (1-simplex) and higher-order (2-simplex) interactions is considered (left panel). Each node follows chaotic dynamics governed by either the R\"{o}ssler or Lorenz system (middle panel). Depending on the coupling strengths, the network exhibits distinct collective behaviors, including desynchronized, synchronized, chimera, and synchronized cluster states. The prevalence and robustness of these states are quantified by varying the pairwise interaction strength ($k_1$) and the higher-order interaction strength ($k_2$) over different initial conditions (right panel). Our primary focus is on understanding how higher-order interactions influence the emergence and robustness of chimera states underlying various interaction mechanisms.} %under three coupling schemes: diffusive, conjugate, and mean-field diffusive coupling.}
    \label{fig:1}
\end{figure} 

The paper is organized as follows. Section~\ref{sec:2} introduces the wheel network topology. Section~\ref{sec:3} formulates the governing equations. Section~\ref{sec:6} defines two new statistical measures that help to categorize resulting dynamical behavior. Section~\ref{subsec:4.1} and \ref{subsec:4.2} presents the dynamical behaviors induced by pairwise and higher-order interactions under different coupling schemes. Section~\ref{subsec:4.3}  develops the analytical framework based on network symmetry, the transverse coupling spectrum, and the master stability function. Finally, Section~\ref{sec:5} concludes with a summary and outlook.
\vspace{-0.5cm}
\section{Methodology}
\subsection{Network Structure}
\label{sec:2}
We consider a wheel network comprising a central hub node connected to all peripheral nodes, while the peripheral nodes are interconnected through nearest-neighbor links (Fig.~\ref{fig:1}). This topology differs from a star network by the presence of these additional peripheral connections, enabling higher-order interactions alongside pairwise couplings. In our study, three-body (higher-order) interactions, modeled as 2-simplices (triangles), are incorporated alongside two-body (pairwise) interactions represented by 1-simplices. 
 In our study, all nodes are governed by identical dynamical equations. Therefore, the peripheral nodes of the wheel network possess identical intrinsic dynamics and coupling environments, and the peripheral ring consequently exhibits rotational symmetry. Consequently, the wheel network provides a minimal yet nontrivial setting for studying chimera states, where the coexistence of synchronized and desynchronized peripheral nodes arises from spontaneous symmetry breaking rather than structural heterogeneity. This makes the wheel topology particularly suitable for investigating the effects of pairwise and higher-order interactions on the emergence and stability of chimera states.

\subsection{Network Dynamics}
\label{sec:3}
We examine the dynamics of a wheel network in which we take dynamics on the nodes to be of three-dimensional chaotic oscillators, specifically the R\"{o}ssler and Lorenz oscillators. The network incorporates both pairwise and higher-order three-body interactions, and its collective behavior is analyzed under three distinct coupling schemes: diffusive, conjugate, and mean-field diffusive coupling. The governing equations for the $i^{th}$ node in the network of coupled oscillators under diffusive coupling are given by
\begin{equation}
  \begin{aligned}
      \dot{x_i} &= f_x(x_i,y_i,z_i) + \frac{k_1}{ d^1_i }\sum_{j=0}^{N-1} A_{ij}(x_j-x_i) \\
      &\quad + \frac{k_2}{2 d^2_i } \sum_{j=0}^{N-1}\sum_{l=0}^{N-1} A_{ijl}(x_j +x_l - 2x_i) \\
      \dot{y_i} &= f_y(x_i,y_i,z_i) \\
      \dot{z_i} &= f_z(x_i,y_i,z_i) 
      .\label{eq:1}
  \end{aligned}
\end{equation}
where \(k_{1}\) and \(k_{2}\) are the strengths of the 1-simplex and 2-simplex interactions, respectively;  \(d_i^{1}\) and \(d_i^{2}\) denote the corresponding degrees of node \(i\); and \(N\) is the network size. In a wheel network, each peripheral node (\(i=1,\ldots,N-1\)) has \(d_i^{1}=3\) and \(d_i^{2}=2\), while the hub node $i=0$ has \(d_0^{1}=d_0^{2}=N-1\).
 1-simplex interactions are encoded by the adjacency matrix $A$, where $A_{ij} = 1$ if $i$ and $j$ are connected, and $A_{ij} = 0$ otherwise. Similarly, in the third term, 2-simplex interactions are encoded by the adjacency tensor matrix $A$, where $A_{ijl}=1$ if nodes $i,j$ and $l$ interact simultaneously and form a triangle, as illustrated in Fig.~\ref{fig:1}, and $A_{ijl}=0$ otherwise. 
 The 2-simplex interaction term represents the collective diffusive influence of the two neighboring nodes forming a 2-simplex with node \(i\). Specifically, the quantity \((x_j+x_l-2x_i)\) measures the deviation of the state of node \(i\) from the collective state of the other two nodes in the 2-simplex and therefore acts as a higher-order generalization of the pairwise diffusive term \((x_j-x_i)\).
 
 Similarly, the dynamical equation for each node $i$ under the conjugate coupling is defined as:
  
\begin{equation}
  \begin{aligned}
      \dot{x_i} &= f_x(x_i,y_i,z_i) 
      + \frac{k_1}{d^1_i }\sum_{j=0}^{N-1} A_{ij}(y_j-x_i) \\
      &\quad + \frac{k_2}{2d^2_i }\sum_{j=0}^{N-1}\sum_{l=0}^{N-1} A_{ijl}(y_j +y_l - 2 x_i)\\
      \dot{y_i} &= f_y(x_i,y_i,z_i) \\
      \dot{z_i} &= f_z(x_i,y_i,z_i)
      \label{eq:2}
  \end{aligned}
\end{equation}

 Underlying this coupling, interactions occur through dissimilar variables. The pairwise coupling term couples the \(y\)-variable of a neighboring node \(j\) to the \(x\)-variable of node \(i\). Likewise, the 2-simplex interaction term captures the collective influence of the \(y\)-variables of the two neighboring nodes \(j\) and \(l\) on the \(x\)-variable of node \(i\), with \(i\), \(j\), and \(l\) forming a triangular 2-simplex.

Under the mean-field diffusive coupling, the equations of each node $i$ are defined as:
\begin{equation}
  \begin{aligned}
      \dot{x_i} &= f_x(x_i,y_i,z_i) + k_1(x_m - x_i)+ k_2(x_{m}^{'} - x_i) \\
      \dot{y_i} &= f_y(x_i,y_i,z_i) \\
      \dot{z_i} &= f_z(x_i,y_i,z_i) 
      .\label{eq:3}
  \end{aligned}
\end{equation}
where \(x_m=\frac{1}{d^1_i } \sum_{j=0}^{N-1} A_{i,j} x_j \) and \(x_{m}^{'}=\frac{1}{2 d^2_i }\sum_{j=0}^{N-1}\sum_{l=0}^{N-1} A_{i,j,l} (x_j + x_l) \). 
\\
Here, \(x_m\) denotes the local mean field arising from the pairwise neighbors of node \(i\), while \(x_m'\) represents the corresponding higher-order local mean field generated by the nodes participating with node \(i\) in its 2-simplex interactions.

In this study, we employ two types of chaotic dynamics, namely modified R\"{o}ssler and Lorenz systems, at each node of the wheel network.
The governing equations for the modified R\"{o}ssler system are given as follows~\cite{nishikawa2009switching}:
\begin{equation}
  \begin{aligned}
      f_x(x_i,y_i,z_i) &= -[w_i + \epsilon(x_i^2 + y_i^2)]y_i - z_i \\
      f_y(x_i,y_i,z_i) &= [w_i + \epsilon(x_i^2 + y_i^2)]x_i + ay_i \\
      f_z(x_i,y_i,z_i) &= b + z_i(x_i- c) 
      .\label{eq:5}
  \end{aligned}
\end{equation}
The parameters are set to \(a = 0.15\), \(b = 0.4\), \(c = 8.5\), \(\omega_i = 0.41\), and \(\varepsilon = 0.0026\), placing the oscillator in its chaotic regime. In this setting, the instantaneous angular velocity of oscillator \(i\) is well approximated by $\omega_i + \varepsilon\,(x_i^2 + y_i^2)$
i.e., a base frequency \(\omega_i\) perturbed by the chaotic amplitude \(x_i^2 + y_i^2\)~\cite{nishikawa2009switching}.

For Lorenz systems \cite{lorenz1963deterministic}, the equations are following:
\begin{equation}
  \begin{aligned}
      f_x(x_i,y_i,z_i) = \sigma(y_i - x_i) \\
      f_y(x_i,y_i,z_i)= (\rho-z_i)x_i - y_i \\
      f_z(x_i,y_i,z_i) = x_iy_i - \beta z_i 
      .\label{eq:6}
  \end{aligned}
\end{equation}

\noindent The Lorenz system exhibits chaotic dynamics for the typical parameter values \(\sigma = 10\), \(\rho = 28\) and \(\beta = 8/3\) so we take this set of parameters in our study.

\subsection{Statistical measures to classify dynamical behaviors}
\label{sec:6}
 In this study, synchronization is characterized based on the $x$-variables of the peripheral nodes, which are structurally and dynamically equivalent under the rotational symmetry of the wheel network.

We find that commonly used statistical measures, such as the Strength of Incoherence (SI) \cite{gopal2014observation, kundu2022higher, wang2024multi}, are inadequate for our purpose because they primarily rely on nearest-neighbor information and therefore fail to detect synchronization between non-adjacent nodes. In situations where nodes located at arbitrary positions in the network synchronize without being nearest   (as indicated in Fig. S1 (a,b) of the supplementary material), or when the primary interest is not spatial synchronization, SI does not reliably identify the underlying synchronization phenomenon.

To overcome this limitation, we introduce two statistical measures designed to capture nonlocal synchronization and to classify the four observed dynamical behaviors. The first measure quantifies the fraction of node pairs (not necessarily node links) that are not synchronized, referred to as the fraction of desynchronized pairs and denoted by $F_{\mathrm{DP}}$, and mathematically, it is defined as

\begin{equation}
    F_{\mathrm{DP}} = \frac{ \sum_{i=1}^{N_{\mathrm{peri}}}\Sigma_{j=1;j \neq i}^{N_{\mathrm{peri}}} \Theta(\langle|x_j - x_i|\rangle_t- \delta)}{N_{\mathrm{peri}}(N_{\mathrm{peri}}-1)}.
    \label{eq:Fdp}
\end{equation}

Here, \(\Theta(\cdot)\) denotes the Heaviside step function and \(\delta=10^{-6}\) is a predefined threshold.  The summations run over all peripheral nodes $(i=1,2,\ldots,N_{\mathrm{peri}}=N-1)$, while the hub node ($i=0$) is excluded from the analysis so that the classification reflects symmetry breaking among the structurally and dynamically equivalent peripheral nodes.

%and mathematically it is defined as:  It is define as ratio of pairs of desynchronized nodes out of the total possible pairs of nodes in the network structure.

%determines the probability of the individually desynchronized nodes out of the total $N$ nodes in the network structure or 
The second measure computes the fraction of nodes that are not synchronized with any other node in the network, referred to as the fraction of desynchronized nodes, and denoted by $F_{\mathrm{DN}}$. Mathematically, it is defined as 
 
\begin{equation}
    F_{\mathrm{DN}} = \frac{\sum_{i=1}^{N_{\mathrm{peri}}}(\Pi_{j=1;j \neq i}^{N_{\mathrm{peri}}}\Theta(\langle |x_j - x_i| \rangle_t - \delta))}{N_{\mathrm{peri}}}.
    \label{eq:Fdn}
\end{equation}

The classification of emergent collective dynamical behaviors depends on the combination of $F_{\mathrm{DP}}$ and $F_{\mathrm{DN}}$ values, as summarized in Table~\ref{tab:tab1}. Our proposed measures successfully capture nonlocal synchronization by comparing the dynamical states of all the peripheral nodes in the network.   Within the definitions adopted in this work, the combinations of $F_{\mathrm{DP}}$ and $F_{\mathrm{DN}}$ exhaust the possible collective states considered in our classification, namely, $(0,0)$ corresponds to complete synchronization, $(1,1)$ to complete desynchronization, $0<F_{\mathrm{DP}},F_{\mathrm{DN}}<1$ to chimera states, and $0<F_{\mathrm{DP}}<1$ with $F_{\mathrm{DN}}=0$ to synchronized cluster states. A detailed interpretation of these measures is provided in Sec.~S1 of the Supplementary Material.

\begin{table}[H]
    \centering
    \begin{tabular}{|c|c|c|}
       \hline
       $\mathbf{F_{\mathrm{DP}}}$ & $\mathbf{F_{\mathrm{DN}}}$ & \textbf{Dynamical Behavior} \\
       \hline
          1 & 1 & Desynchronization \\ \hline
          0 & 0 & Synchronization \\ \hline
          $0<F_{\mathrm{DP}}<1$ & $0<F_{\mathrm{DN}}<1$ &Chimera  \\ \hline
         $0<F_{\mathrm{DP}}<1$ & 0 & Synchronized Clusters \\ 
         \hline
    \end{tabular}
    \caption{Classification of collective dynamical behaviors using the statistical measures $F_{\mathrm{DP}}$ and $F_{\mathrm{DN}}$.}
    \label{tab:tab1}
\end{table}
The primary objective of this study is to elucidate the influence of higher-order interactions on the emergence and robustness of chimera states across different coupling mechanisms and oscillator dynamics, and to determine whether such interactions promote or suppress chimera behavior.  To address this objective, we perform numerical simulations of coupled chaotic oscillators under all coupling schemes and characterize the collective behavior of the peripheral nodes through the $x$ variable, since the coupling acts only through this variable. The hub node is excluded from the state classification to focus on symmetry breaking among the peripheral nodes. Nevertheless, it remains an active part of the network dynamics and can mediate synchronization between peripheral nodes. Initially, we consider pairwise interactions, and subsequently extend the framework to incorporate higher-order interactions, specifically considering three-body interactions.

\vspace{-\baselineskip}
%\newpage
\section{RESULTS}
\label{sec:4}
First, we examine the collective dynamical behaviors of the wheel network by performing numerical simulations for selected combinations of coupling strengths $(k_1, k_2)$. Further, to determine the robustness of dynamical behaviors, we run simulations over multiple initial conditions and compute the fraction of initial conditions that lead to each dynamical regime.  In addition, using proposed statistical measures, we classify dynamical regimes in the ($k_1, k_2$) parameter space. The following subsections present detailed insights into how higher-order interactions influence the emergence, prevalence, and robustness of these collective behaviors, with particular emphasis on chimera states.

 Throughout the Results section, synchronized, desynchronized, chimera, and synchronized cluster states are denoted by green, blue, red, and magenta colors, respectively. Numerical simulations are performed using the fourth-order Runge--Kutta method with a time step of \(dt=0.01\). To eliminate transient effects, the first \(7\times10^{5}\) integration points are discarded, and the subsequent $100$ time points of the peripheral node time series are used to characterize the dynamical behavior. The initial conditions for all variables, \(x_i\), \(y_i\), and \(z_i\), are chosen independently from a uniform distribution over the interval \([-2,2]\).

\subsection{Chimera states in wheel network of coupled chaotic oscillators}
\label{subsec:4.1}

\begin{figure}[t]
    \centering
    \includegraphics[width=0.49\linewidth]{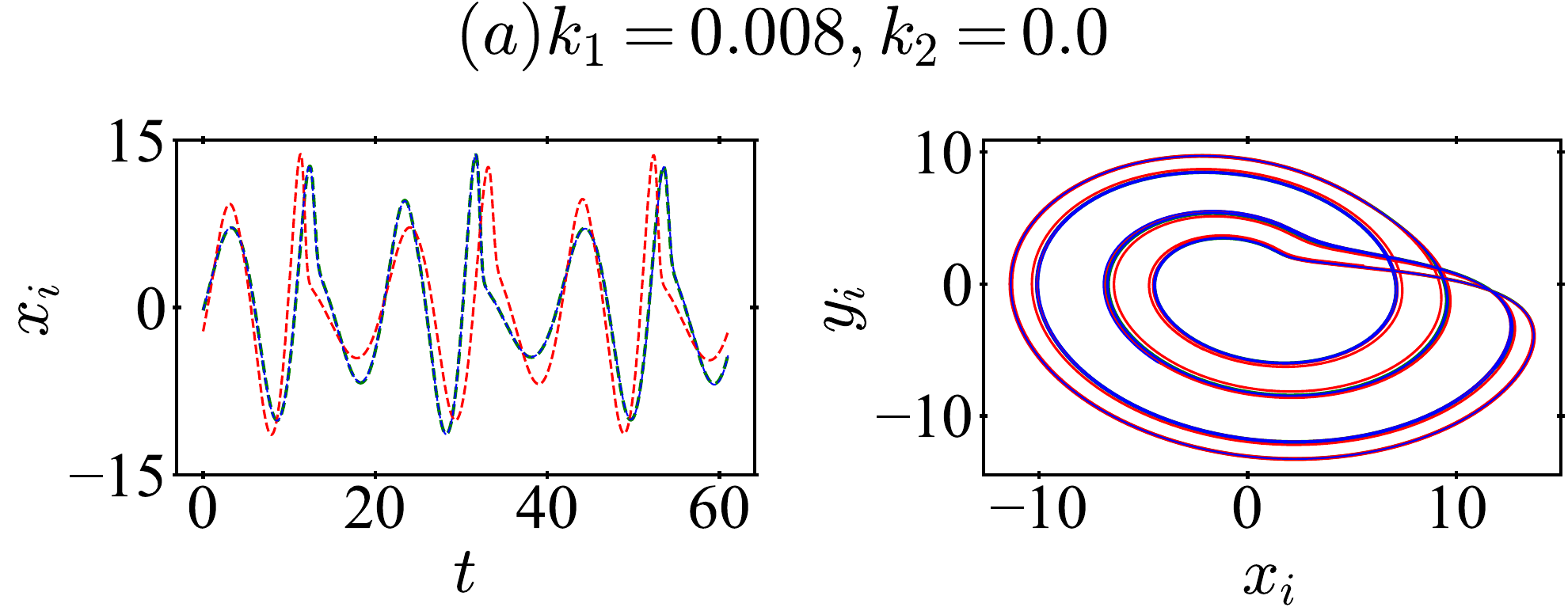} \includegraphics[width=0.49\linewidth]{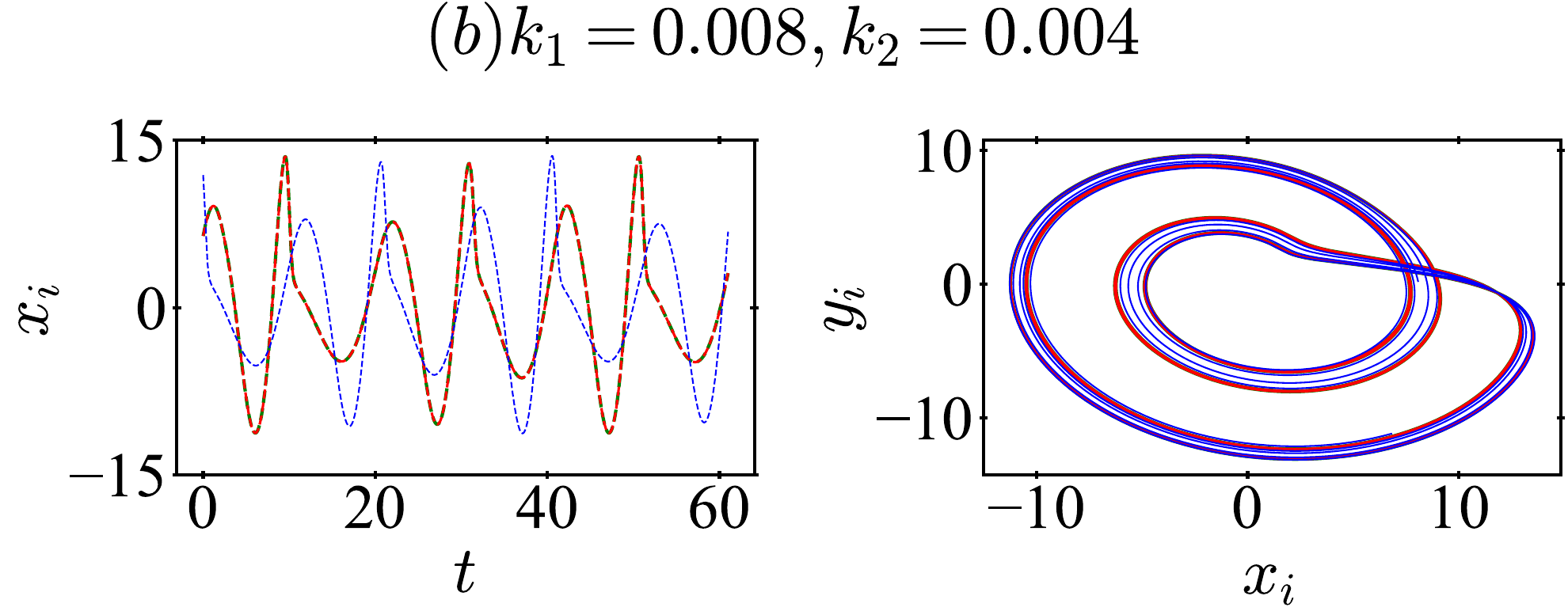}\\
    \includegraphics[width=0.49\linewidth]{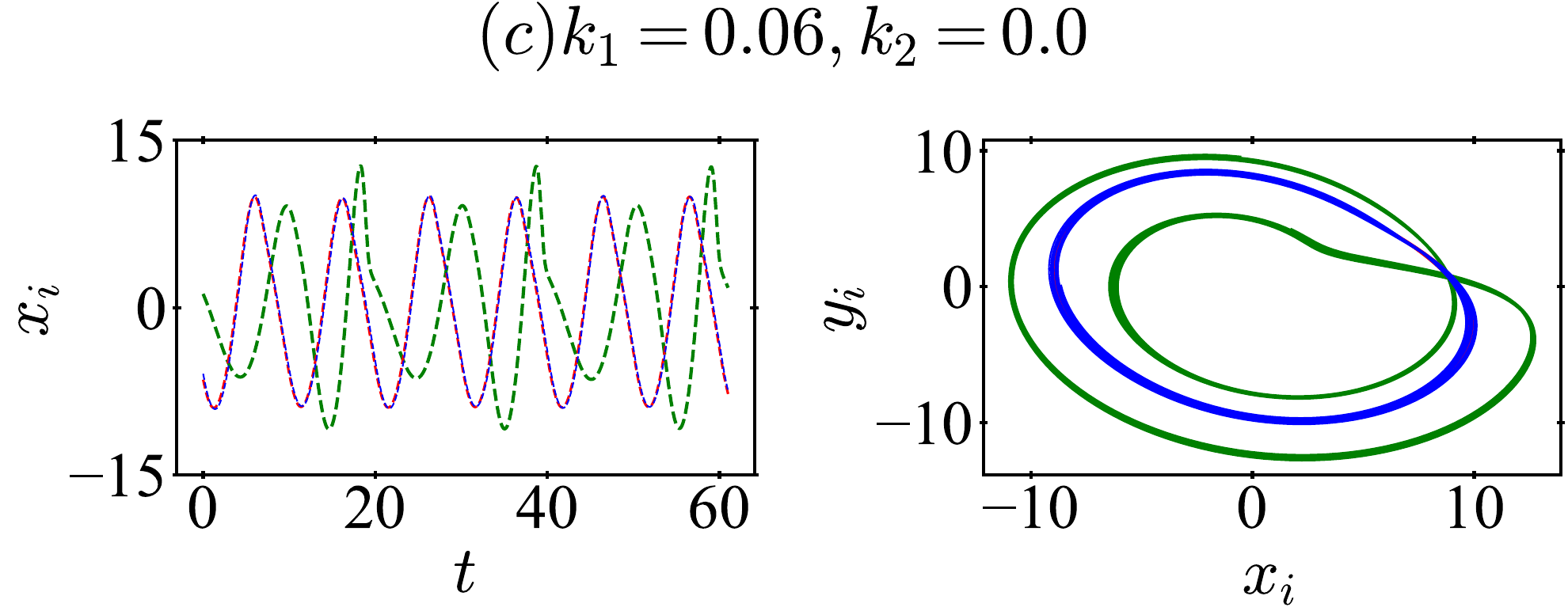} \includegraphics[width=0.49\linewidth]{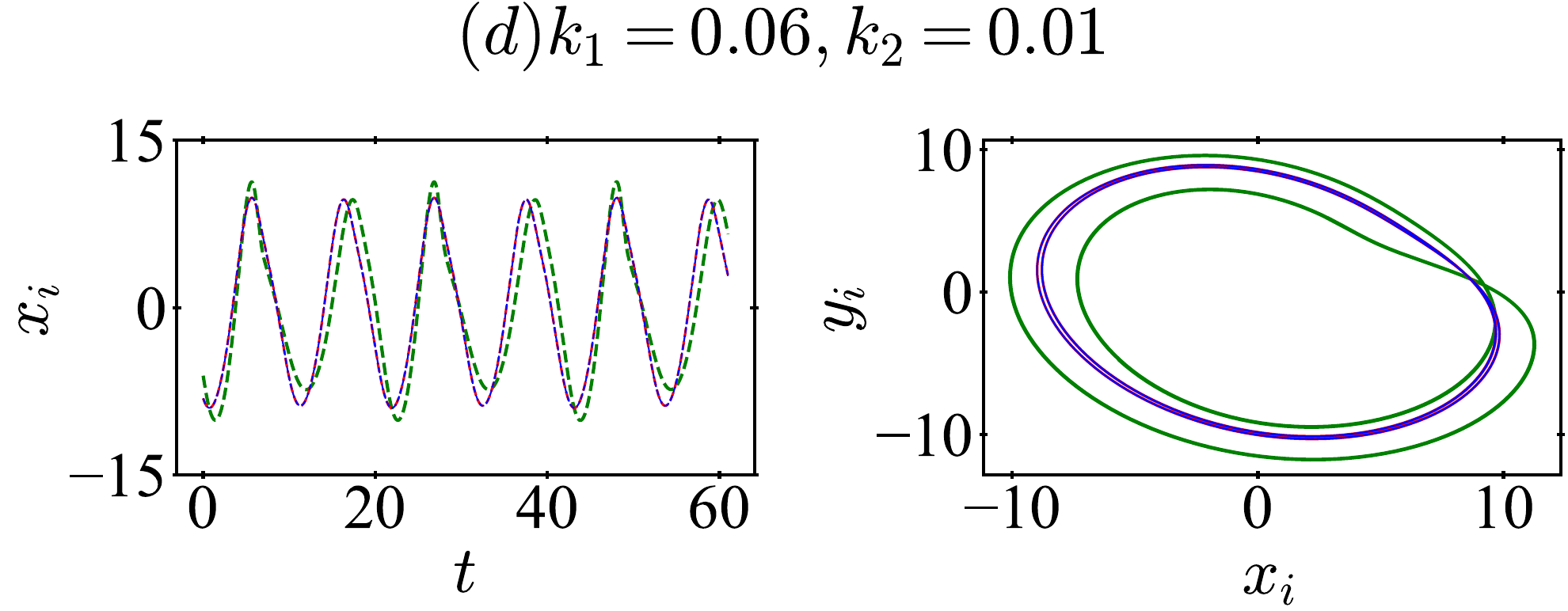}\\
    \includegraphics[width=0.49\linewidth]{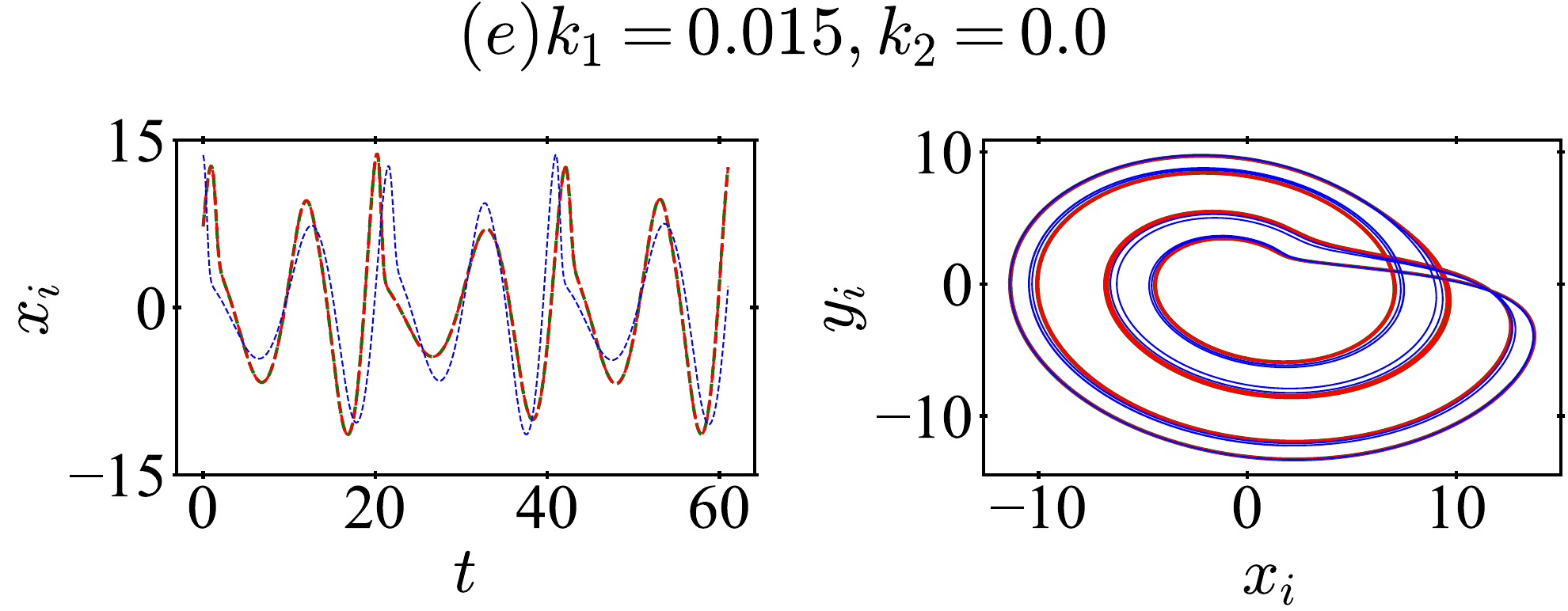}
    \includegraphics[width=0.49\linewidth]{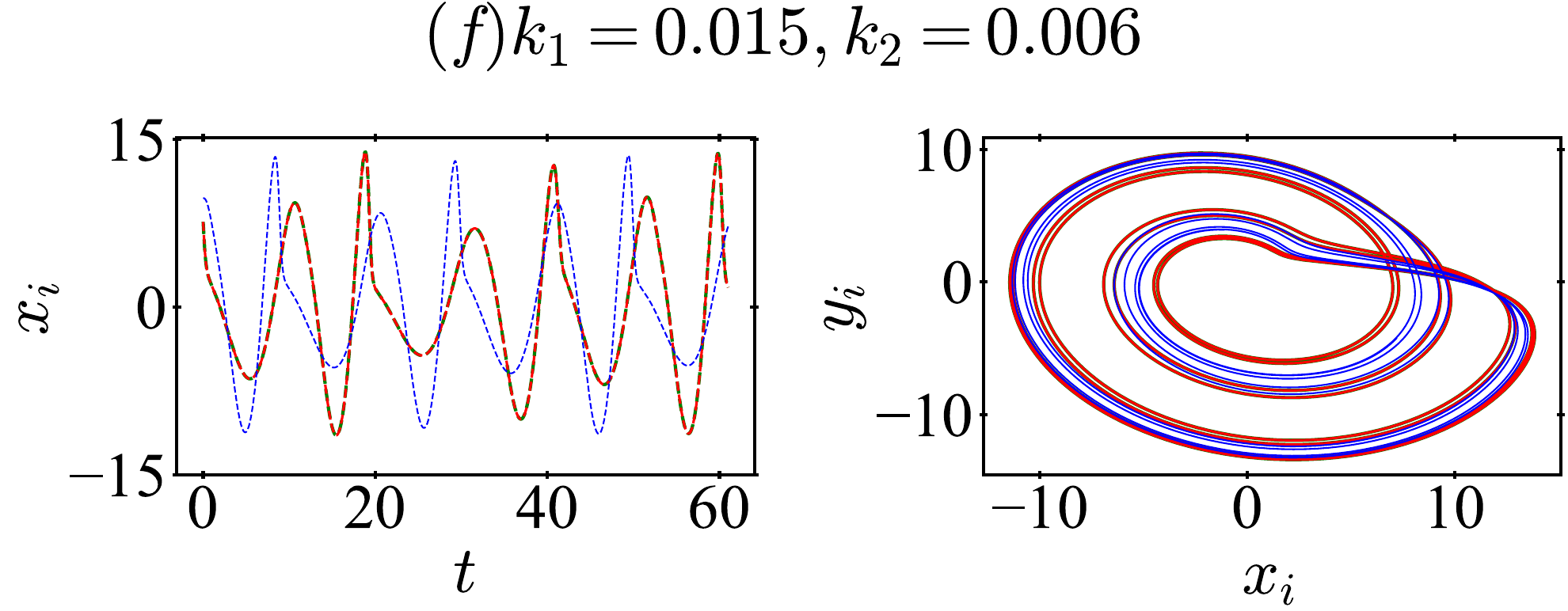}
   \caption{\textbf{Representative time series and phase portraits for a wheel network of four coupled R\"{o}ssler oscillators.} Time series and phase portraits of the three peripheral nodes under three coupling schemes are shown.  Panels (a,b) correspond to diffusive coupling with $(k_1,k_2)=(0.008,0)$ and $(0.008,0.004)$, respectively. Panels (c,d) correspond to conjugate coupling with $(k_1,k_2)=(0.075,0)$ and $(0.06,0.01)$, respectively. Panels (e,f) correspond to mean-field diffusive coupling with $(k_1,k_2)=(0.02,0)$ and $(0.015,0.006)$, respectively. For all coupling schemes, one of the three peripheral nodes remains desynchronized while the other two synchronize, resulting in a chimera state with $F_{DP}=0.67$ and $F_{DN}=0.33$.}
     \label{fig:2}
    
\end{figure}

\begin{figure}[t]
    \centering
    \includegraphics[width=0.49\linewidth]{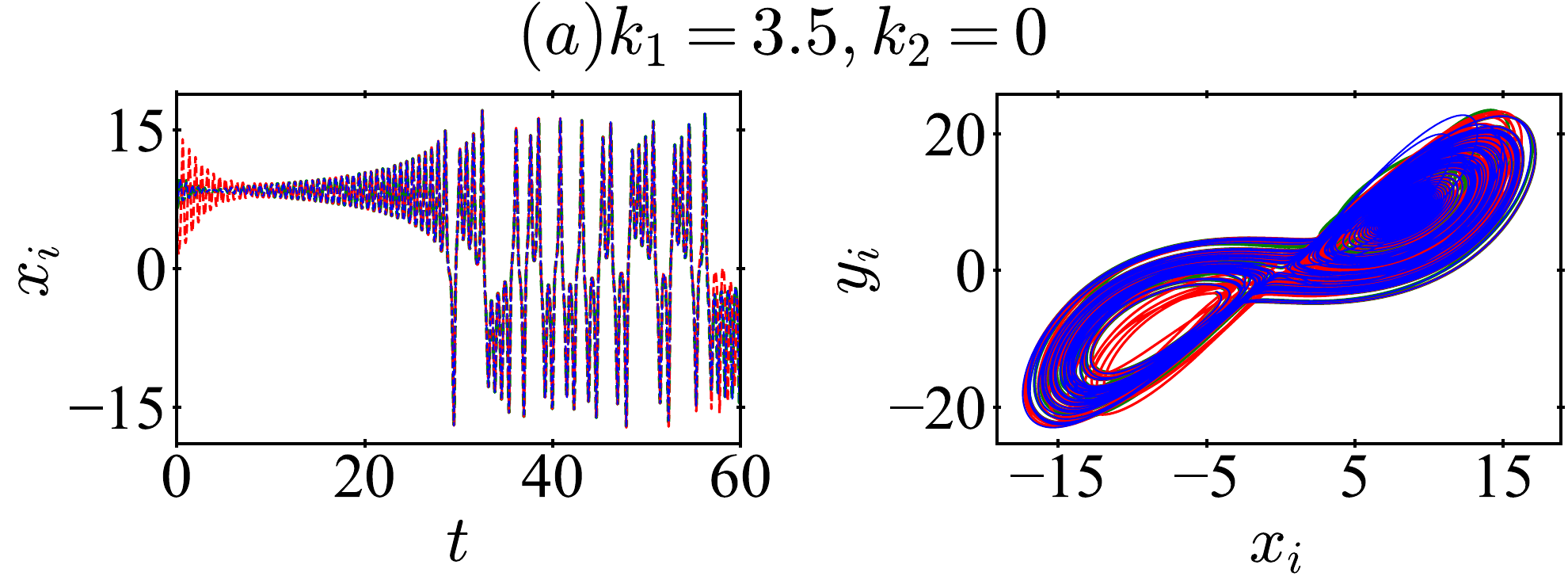} \includegraphics[width=0.49\linewidth]{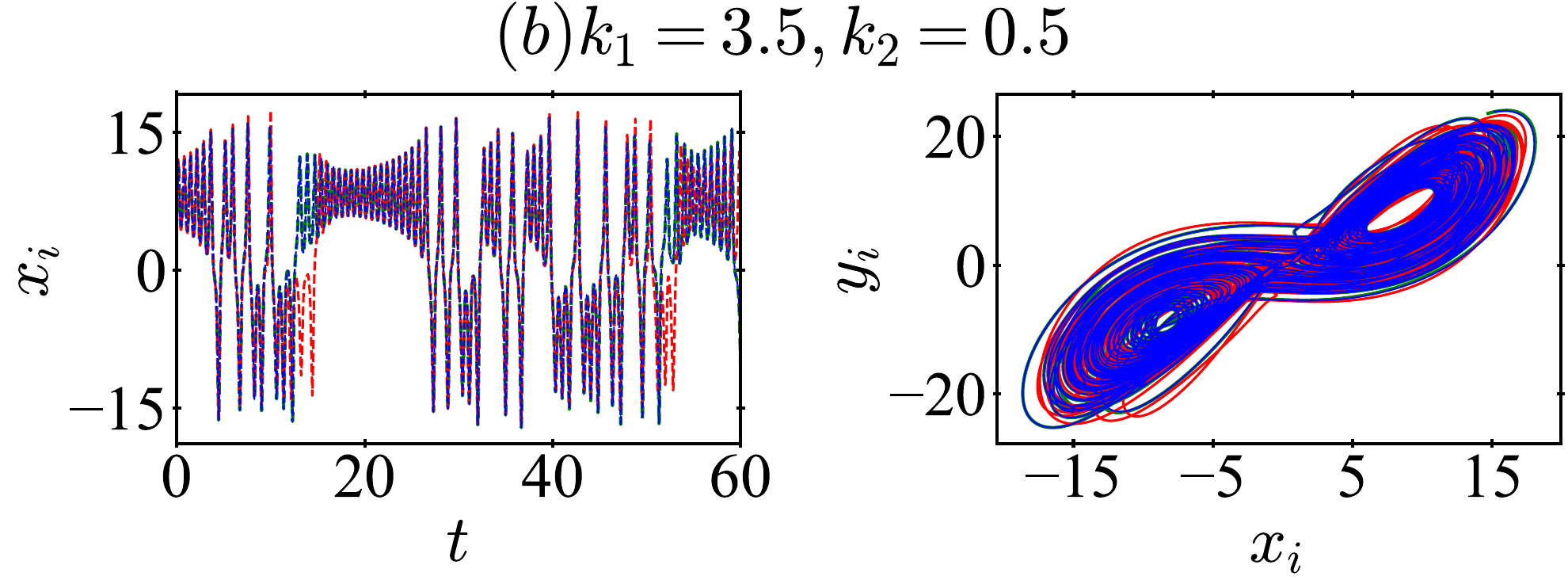}\\
    \includegraphics[width=0.49\linewidth]{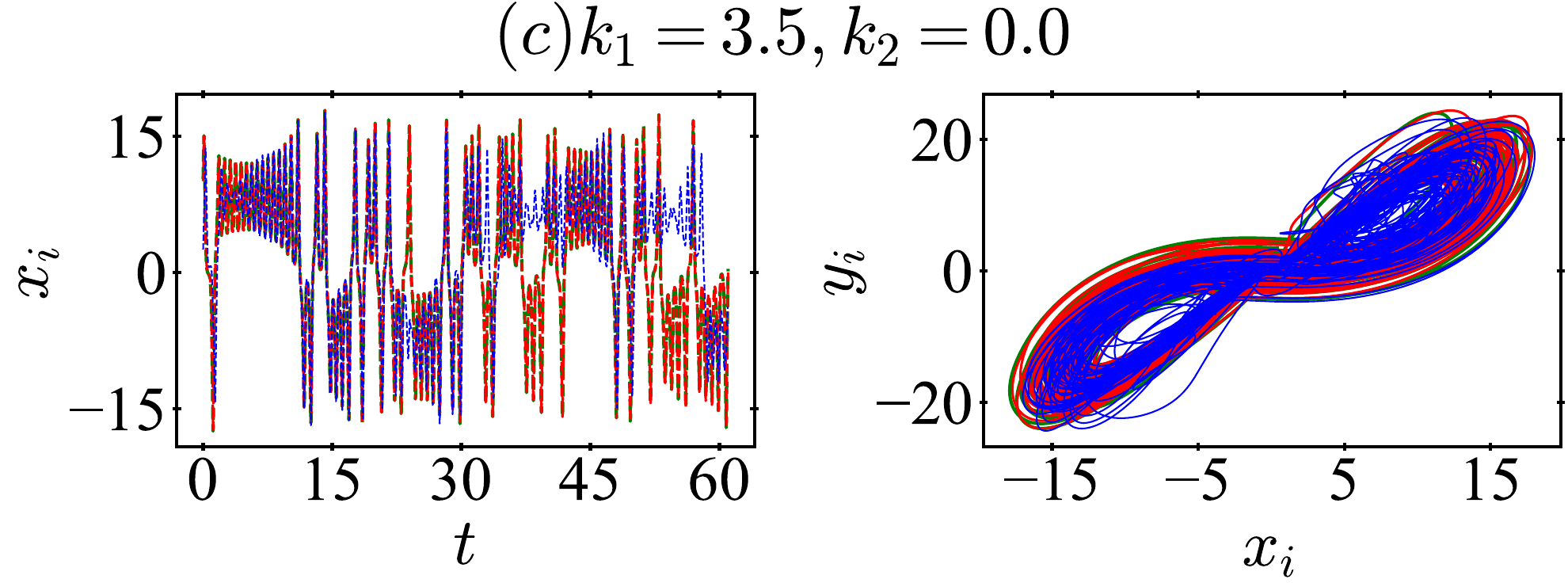} \includegraphics[width=0.49\linewidth]{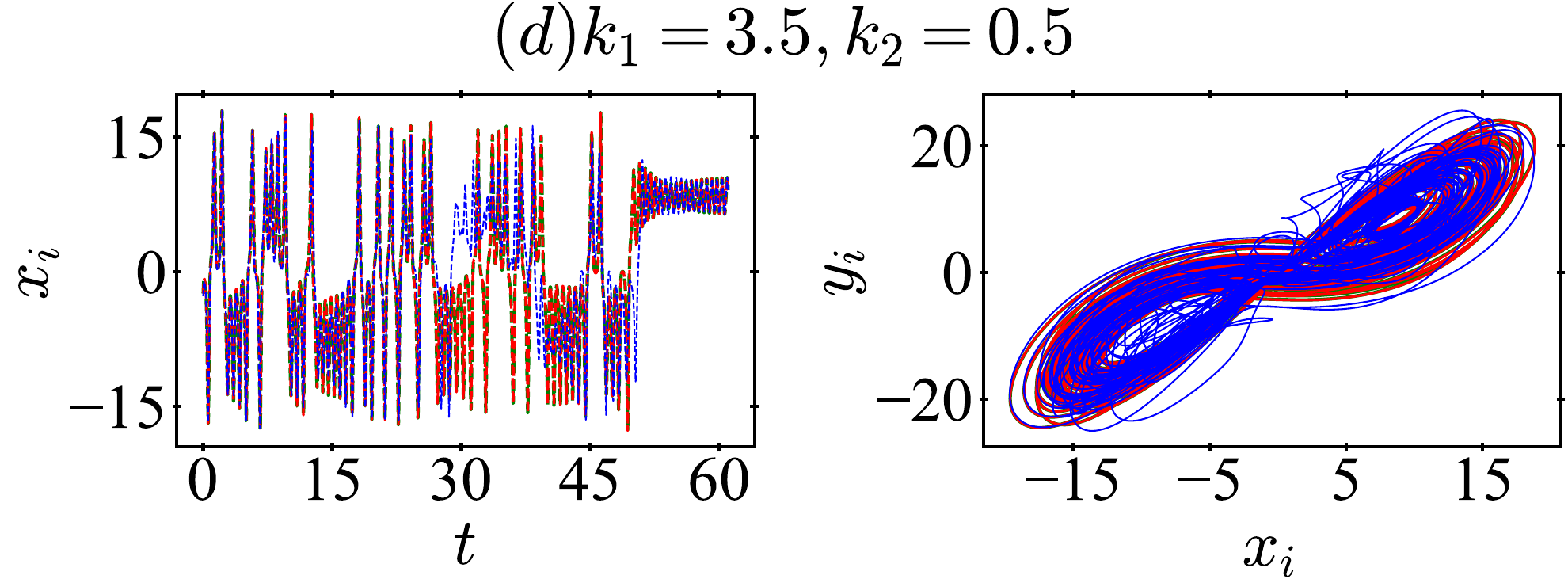}\\
    \includegraphics[width=0.49\linewidth]{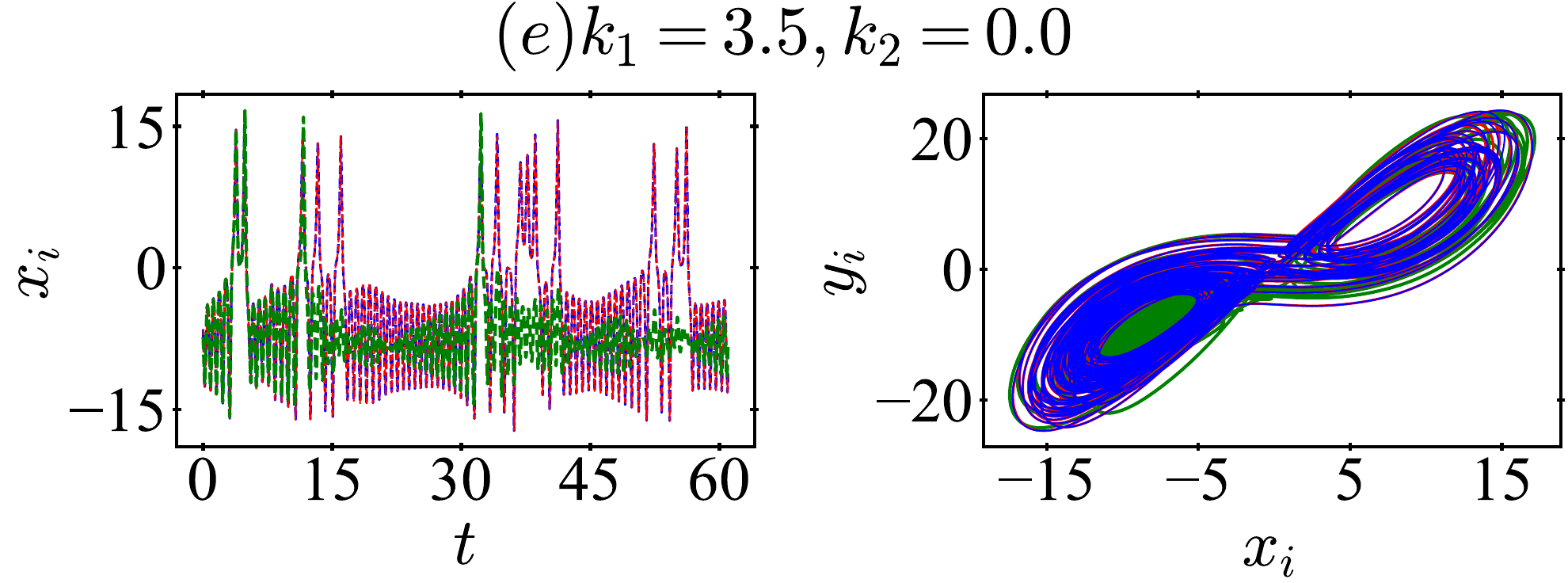}
    \includegraphics[width=0.49\linewidth]{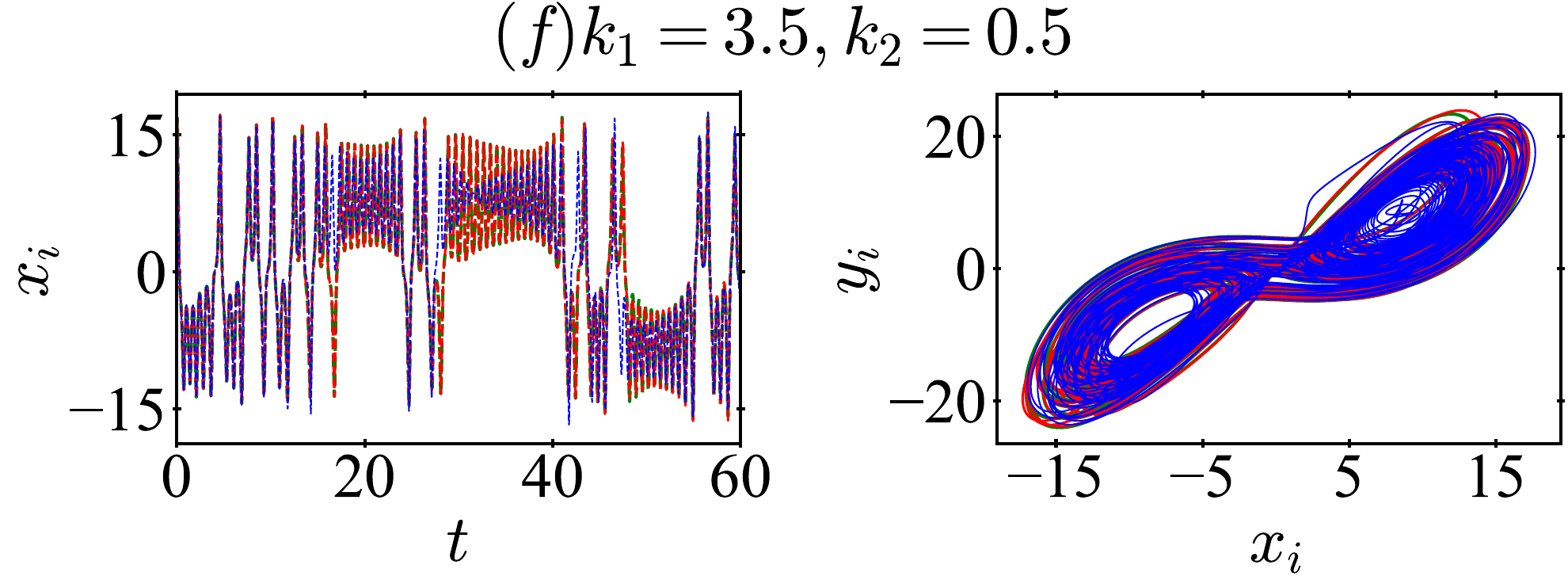}
   \caption{\textbf{Representative time series and phase portraits for a wheel network of four coupled Lorenz oscillators.} All panels display the time series and phase portraits of the three peripheral nodes only. For each coupling scheme, the left panel shows the case with pairwise interactions only, while the right panel shows the case with both pairwise and higher-order interactions.  For each coupling scheme, the left panel shows the case with pairwise interactions only $(k_1=3.5,\; k_2=0)$, while the right panel shows the case with both pairwise and higher-order interactions $(k_1=3.5,\; k_2=0.5)$. For all the cases depicted above, two peripheral nodes evolve synchronously while the third remains desynchronized, with all of these dynamics resulting into $F_{DP} = 0.67$ and $F_{DN}= 0.33$, indicating a chimera state.}
     \label{fig:3}
    
\end{figure}

 Previous studies have demonstrated that chimera states can also occur in networks comprising only a few oscillators~\cite{wojewoda2016smallest,ashwin2015weak}. Motivated by these observations, we first examine a wheel network of size $N=4$ for coupled R\"{o}ssler  and Lorenz oscillators, as shown in Figs.~\ref{fig:2} and \ref{fig:3}, respectively. The small network size is chosen primarily for visualization, since the dynamics of the three peripheral nodes can be observe clearly from their time series and corresponding phase portraits. Results are presented for all three coupling mechanisms considered here (cf. Figs.~\ref{fig:2} and \ref{fig:3}) with the left and right panels corresponding to systems with pairwise interactions alone and with both pairwise and higher-order interactions, respectively.

For a wheel network with $N=4$, there are three peripheral nodes, of which two evolve synchronously while the remaining node exhibits distinct dynamics. This behavior is clearly observed in the representative cases shown in Figs.~\ref{fig:2} and \ref{fig:3} for both R\"{o}ssler and Lorenz oscillators under all three coupling mechanisms. Thus, despite being structurally and dynamically equivalent, the peripheral nodes spontaneously separate into coherent and incoherent subsets, providing a direct manifestation of dynamical symmetry breaking and the characteristic coexistence associated with a chimera state.

 The visual identification is further supported by the statistical measures $F_{\mathrm{DP}}$ and $F_{\mathrm{DN}}$. Figs.~\ref{fig:2} and \ref{fig:3} show that two of the three peripheral nodes are synchronized, giving $F_{\mathrm{DP}}=0.67$, while one of the them belongs to the desynchronized subset, giving $F_{\mathrm{DN}}=0.33$. Consequently, both measures satisfy $0<F_{\mathrm{DP}},F_{\mathrm{DN}}<1$ consistent with our criterion for identifying a chimera state. The agreement between these quantitative measures and the time series and phase-portrait observations therefore provides a consistent classification of the observed states as chimera states for both oscillator models and across the different coupling mechanisms considered here.
 
 Further, to verify observed chimera states, we consider a larger wheel network with $N=100$ oscillators. Representative spatiotemporal plots and corresponding snapshots of the $x$ variable are presented in Fig.~S1 of the Supplementary Material, where the coexistence of synchronized and desynchronized groups among the peripheral nodes is again clearly observed. 
%For such larger networks, where visual identification through time series and phase-portrait becomes less straightforward, the measures $F_{\mathrm{DP}}$ and $F_{\mathrm{DN}}$ provide a systematic quantitative classification of the collective dynamical states. 
We further note that the particular partition of the peripheral nodes into coherent and incoherent groups is not fixed, but may depend on the intrinsic oscillator dynamics, initial conditions, coupling mechanism, and coupling strengths.

\subsection{Prevalence of chimera states}
\label{subsec:4.2}
%To check the prevalence of synchronization, desynchronization, chimera, and synchronized cluster states.
%We observe synchronization, desynchronization, chimera, and synchronized cluster states \cite{starnetwork} in a coupled network depending on initial conditions and $k_1$ and $k_2$ values. 

\begin{figure}[t]
    \centering
    \setlength{\tabcolsep}{1pt}
    \renewcommand{\arraystretch}{0.85}

    \begin{tabular}{@{} m{0.0\columnwidth} c c c @{}}
        & \fontsize{6pt}{8pt}\selectfont \textbf{Diffusive} & \fontsize{6pt}{8pt}\selectfont \textbf{Conjugate} & \fontsize{6pt}{8pt}\selectfont \textbf{Mean-Field Diffusive} \\

        & 
        \parbox[c]{0.32\columnwidth}{\centering\tiny (a)$\,
        {\scriptstyle k_2 = 0}$\\ \includegraphics[width=\linewidth]{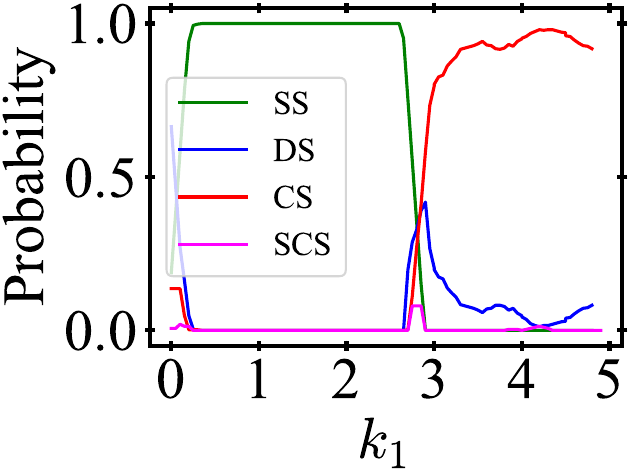}} &
        \parbox[c]{0.32\columnwidth}{\centering \tiny (b)$\, {\scriptstyle k_2 = 0}$ \\ \includegraphics[width=\linewidth]{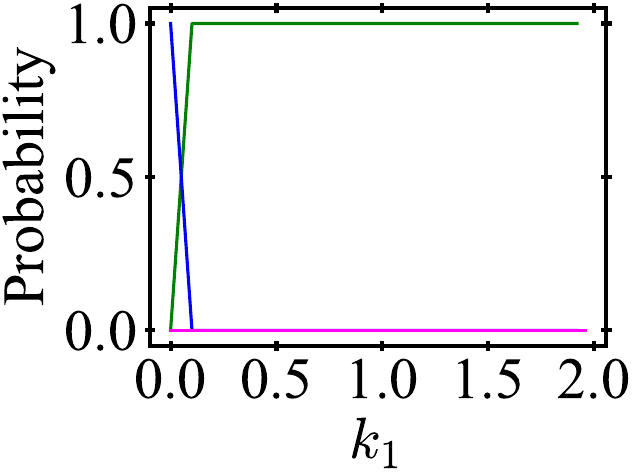}} &
        \parbox[c]{0.32\columnwidth}{\centering \tiny (c)$\, {\scriptstyle k_2 = 0}$ \\ \includegraphics[width=\linewidth]{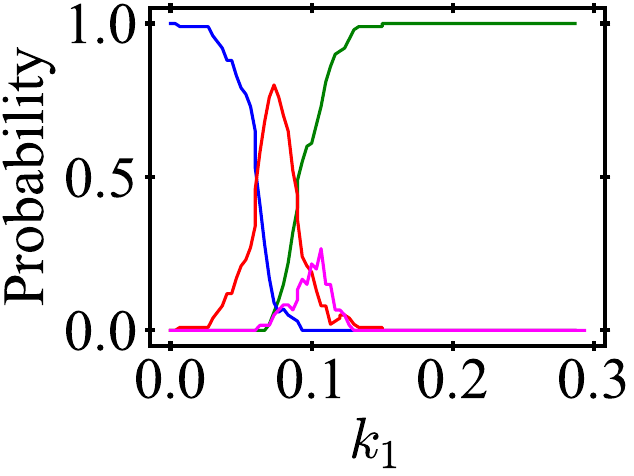}} \\

        & 
        \parbox[c]{0.32\columnwidth}{\centering \tiny (d)$\, {\scriptstyle k_{2} = 2}$ \\ \includegraphics[width=\linewidth]{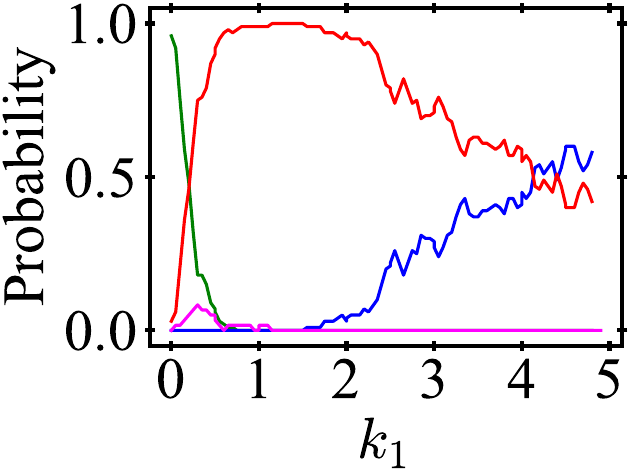}} &
        \parbox[c]{0.32\columnwidth}{\centering \tiny (e)$\, {\scriptstyle k_{2} = 1.01}$ \\ \includegraphics[width=\linewidth]{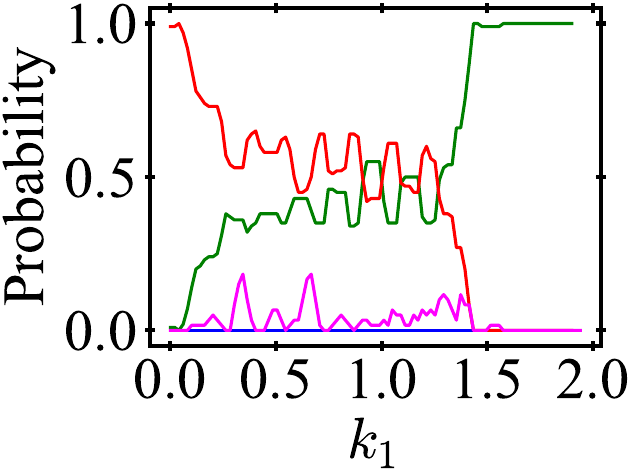}} &
        \parbox[c]{0.32\columnwidth}{\centering \tiny (f)$\, {\scriptstyle k_{2} = 0.05}$ \\ \includegraphics[width=\linewidth]{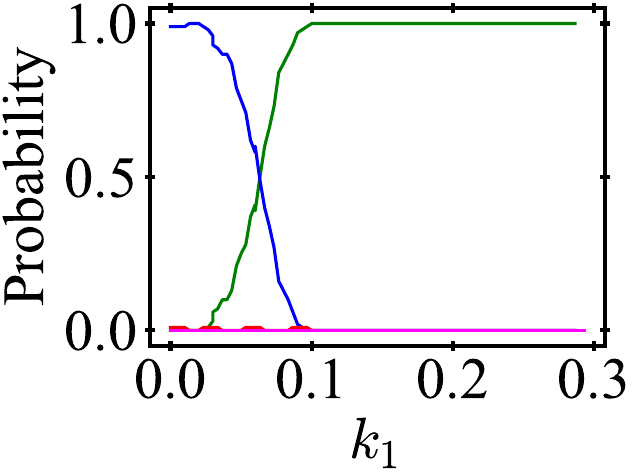}} \\
    \end{tabular}

    \caption{\textbf{Probabilities of dynamical behaviors in a wheel network of coupled R\"{o}ssler oscillators.} The probabilities of observing chimera state (CS) in red, synchronized cluster state (SCS) in magenta, synchronized state (SS) in green, and desynchronized state (DS) in blue, for a wheel network of $N=100$ coupled R\"{o}ssler oscillators are shown as functions of the pairwise interaction strength $k_1$. Panels (a--c) correspond to pairwise interactions only ($k_2=0$), while panels (d--f) show the effect of higher-order interactions, with the 2-simplex interaction strengths chosen as  $k_2=2$, $2$, $1.01$, and $0.05$ for the diffusive, conjugate, and mean-field diffusive coupling schemes, respectively.}
    \label{fig:4}
\end{figure}

\begin{figure}[t]
    \centering
    \setlength{\tabcolsep}{1pt}
    \renewcommand{\arraystretch}{0.85}

    \begin{tabular}{@{} m{0.00\columnwidth} c c c @{}}
      & \fontsize{6pt}{8pt}\selectfont \textbf{Diffusive} & \fontsize{6pt}{8pt}\selectfont \textbf{Conjugate} & \fontsize{6pt}{8pt}\selectfont \textbf{Mean-Field Diffusive} \\

        & 
        \parbox[c]{0.32\columnwidth}{\centering\tiny (a)$\, {\scriptstyle k_2 = 0}$\\ \includegraphics[width=\linewidth]{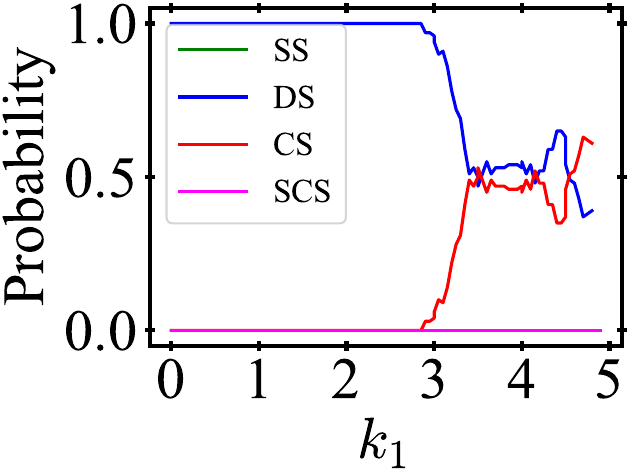}} &
        \parbox[c]{0.32\columnwidth}{\centering \tiny (b)$\, {\scriptstyle k_2 = 0}$ \\ \includegraphics[width=\linewidth]{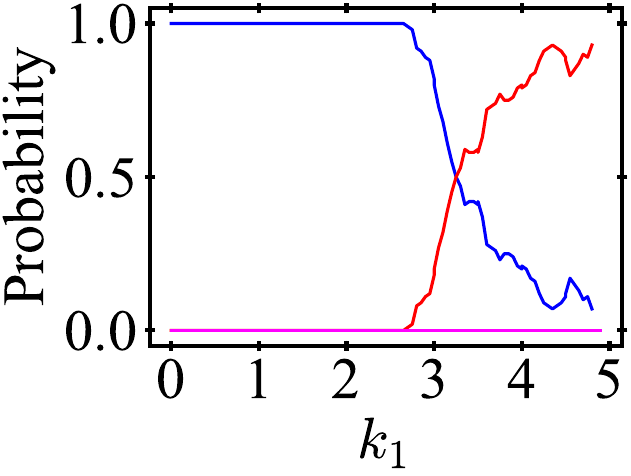}} &
        \parbox[c]{0.32\columnwidth}{\centering \tiny (c)$\, {\scriptstyle k_2 = 0}$ \\ \includegraphics[width=\linewidth]{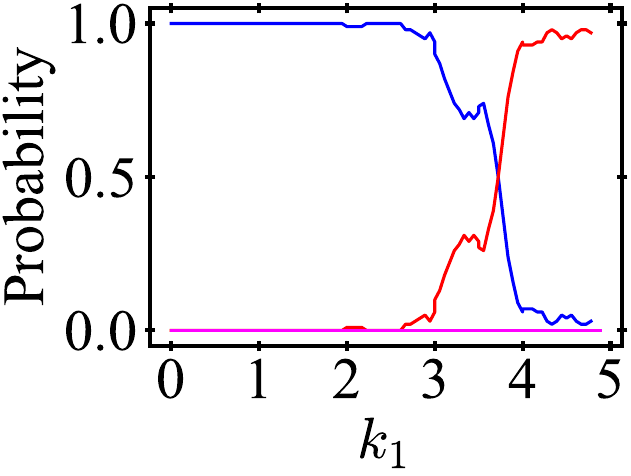}} \\

        & 
        \parbox[c]{0.32\columnwidth}{\centering \tiny (d)$\, {\scriptstyle k_2 = 2}$ \\ \includegraphics[width=\linewidth]{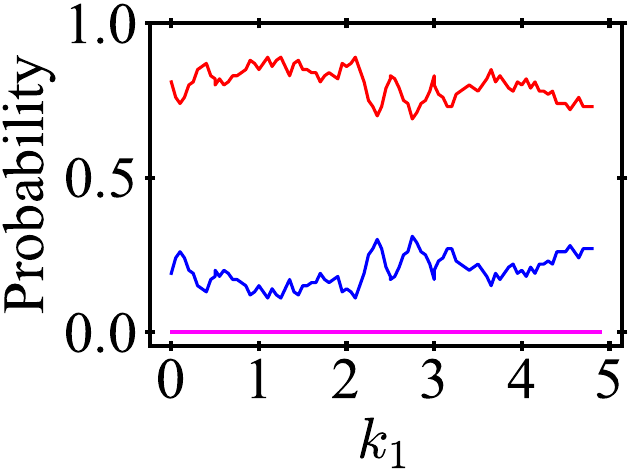}} &
        \parbox[c]{0.32\columnwidth}{\centering \tiny (e)$\, {\scriptstyle k_2 = 2.5}$ \\ \includegraphics[width=\linewidth]{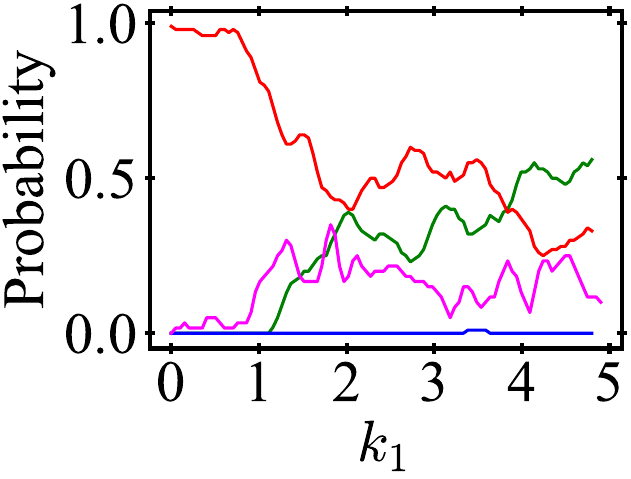}} &
        \parbox[c]{0.32\columnwidth}{\centering \tiny (f)$\, {\scriptstyle k_2 = 2}$ \\ \includegraphics[width=\linewidth]{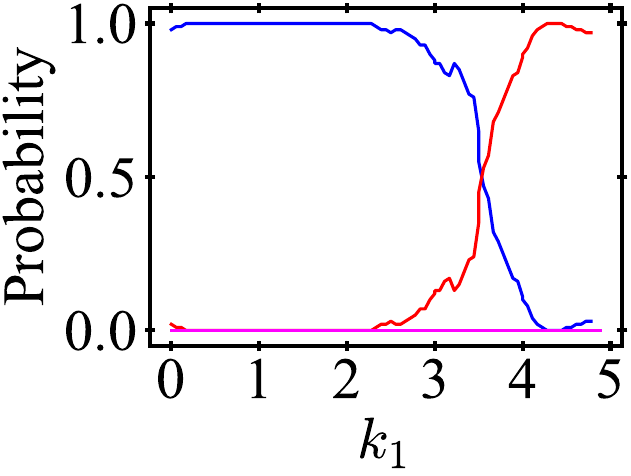}}\\
    \end{tabular}

    \caption{\textbf{Probabilities of different dynamical states in a wheel network of coupled Lorenz oscillators.} The probabilities of observing chimera state (CS) in red, synchronized cluster state (SCS) in magenta, synchronized state (SS) in green, and desynchronized state (DS) in blue, for a wheel network of $N=100$ coupled Lorenz oscillators. Panels (a,d), (b,e), and (c,f) correspond to diffusive, conjugate, and mean-field diffusive coupling, respectively. Panels (a--c) show the results for pairwise interactions only ($k_2=0$),  while panels (d--f) show the results in the presence of higher-order interactions with fixed $k_2=2$, $2.5$, and $2$, respectively.}
    \label{fig:5}
\end{figure}

To quantify the likelihood of observing four dynamical states, we perform simulations for  $100$ random initial conditions, with all state variables initialized independently from a uniform distribution over the interval \([-2,2]\). First, for a fixed value of $k_2$, we vary $k_1$ and compute the fraction of realizations that evolve into each dynamical regime. This fraction provides a numerical estimate of the relative basin size associated with a given state and therefore reflects its prevalence and robustness in the coupled R\"{o}ssler and Lorenz systems, as shown in Figs.~\ref{fig:4} and \ref{fig:5}, respectively.

\begin{figure}[t]
    \centering
    \includegraphics[width=0.49\linewidth]{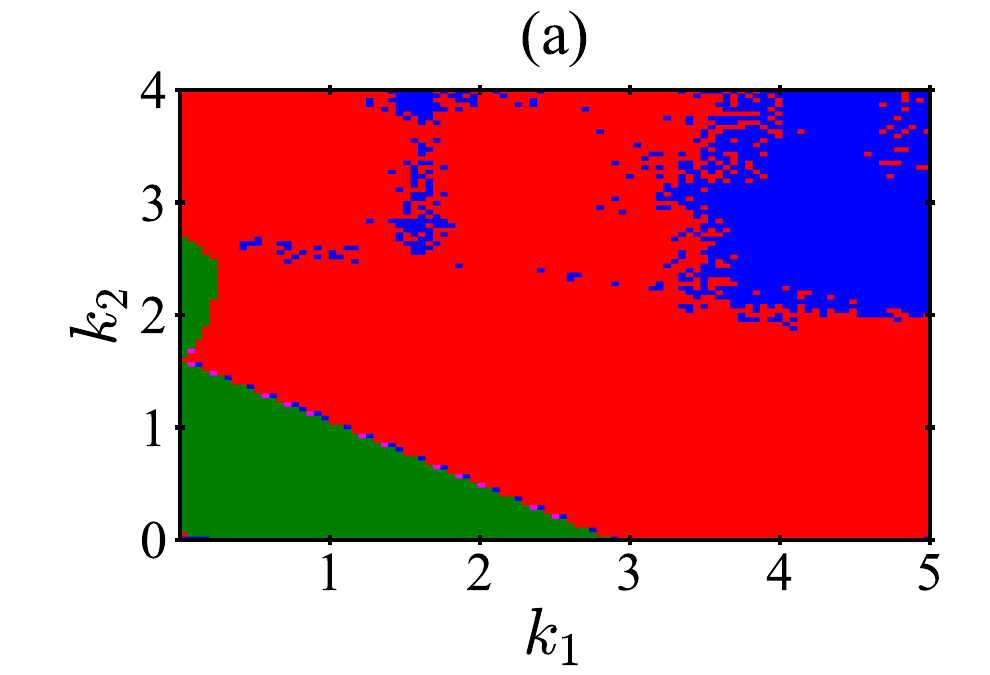} \includegraphics[width=0.49\linewidth]{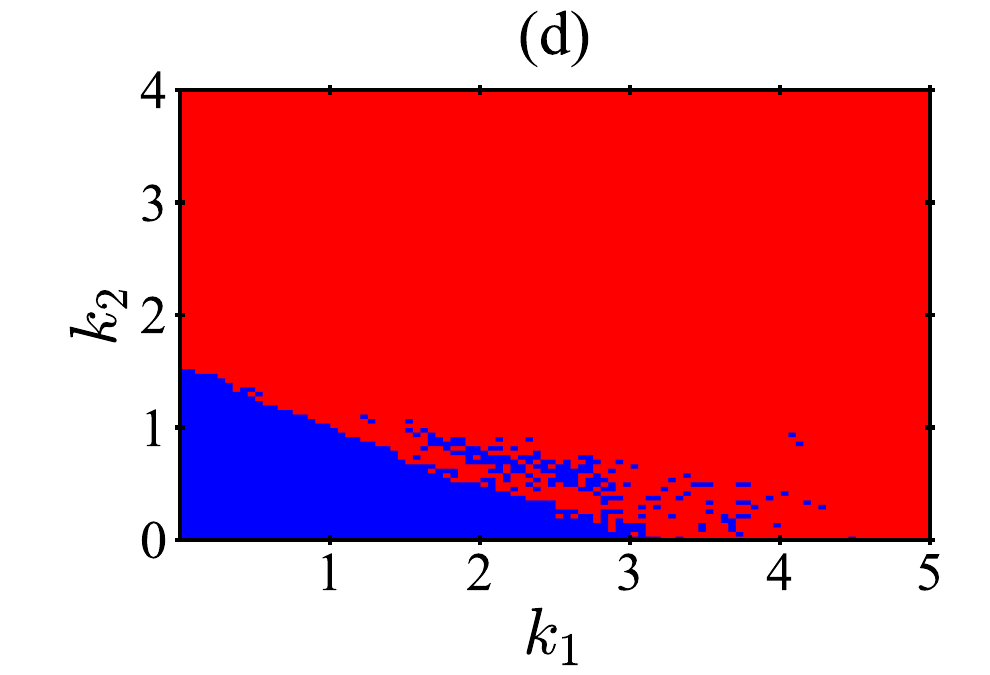}\\
    \includegraphics[width=0.49\linewidth]{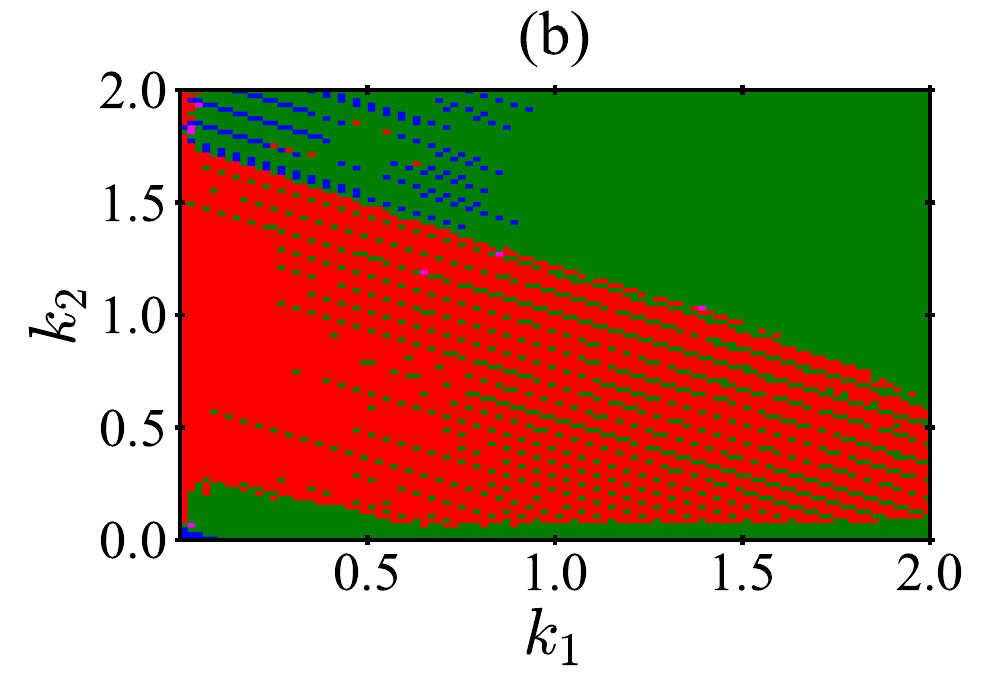} \includegraphics[width=0.49\linewidth]{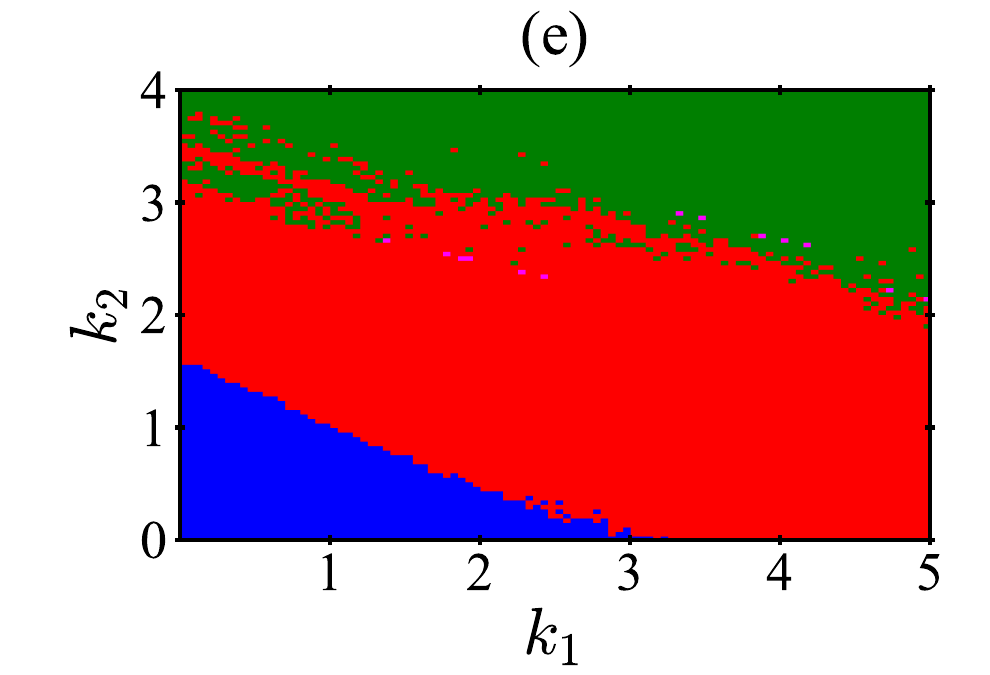}\\
    \includegraphics[width=0.49\linewidth]{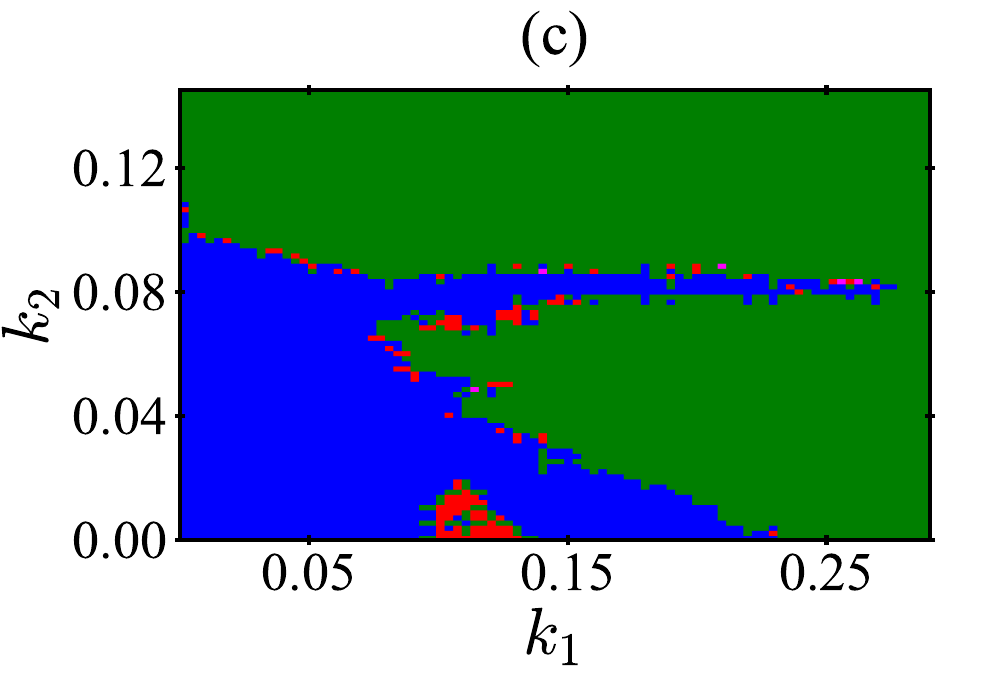}
    \includegraphics[width=0.49\linewidth]{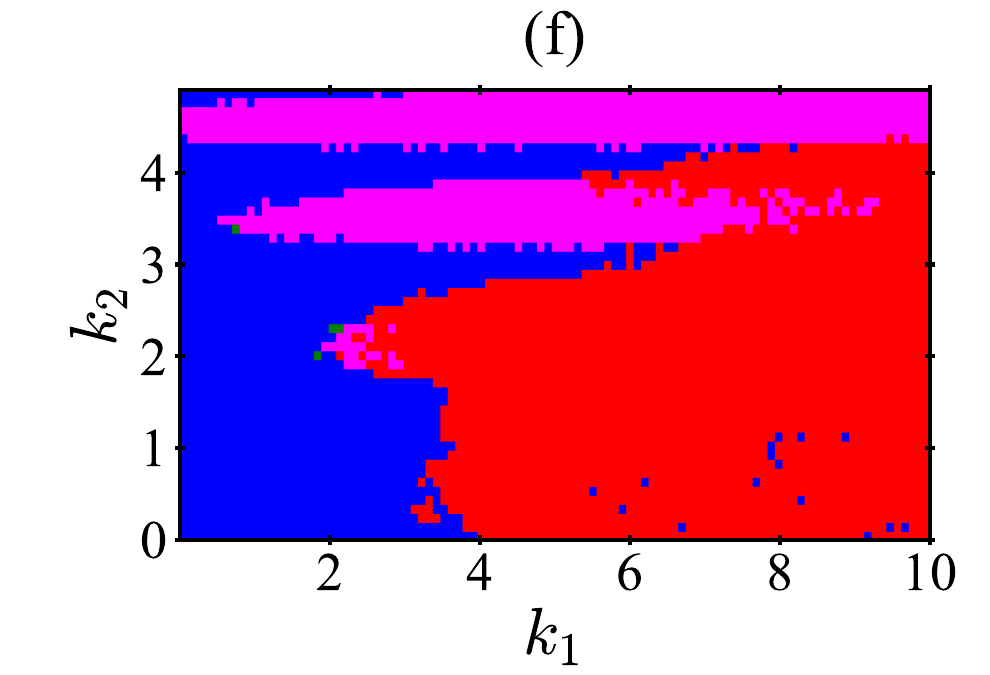}\\
    \includegraphics[width=\linewidth]{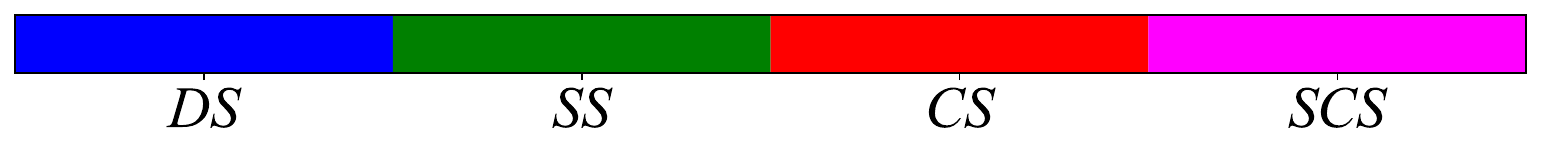}
   \caption{\textbf{Dynamical regimes in the $(k_1,k_2)$ parameter space for a wheel network of $N=100$ coupled chaotic oscillators.} The most probable dynamical state is shown for coupled R\"{o}ssler oscillators in panels (a--c) and coupled Lorenz oscillators in panels (d--f) under three different coupling schemes. Panels (a,d), (b,e), and (c,f) correspond to diffusive, conjugate, and mean-field diffusive coupling , respectively, as described by Eqs.~(\ref{eq:1}–\ref{eq:3}). Blue, green, red, and magenta regions denote desynchronized (DS), synchronized (SS), chimera (CS), and synchronized cluster (SCS) states, respectively.}
    \label{fig:6}
    
\end{figure}

Furthermore, to investigate the regime of the four dynamical states, we systematically vary the coupling strengths $k_1$ and $k_2$ and perform simulations for $100$ independent random initial conditions at each ($k_1, k_2$) pair. The state occurring with the highest probability is identified as the dominant dynamical regime and is used to construct the phase diagrams shown in Fig.~\ref{fig:6}.

 \textbf{Diffusively coupled chaotic oscillators:}
By comparing Figs.~\ref{fig:4}(a) and \ref{fig:4}(d) for $N=100$ diffusively coupled R\"{o}ssler oscillators, we observe that, in the absence of higher-order interactions, chimera states occur predominantly at relatively large values of the pairwise coupling strength $k_1$. Upon introducing higher-order interactions, the chimera region shifts toward smaller $k_1$. This trend is also evident from the $(k_1,k_2)$ parameter space in Fig.~\ref{fig:6}(a) and is consistent with previous studies~\cite{starnetwork,acharyya2011desynchronization}. A similar shift toward smaller $k_1$ is observed for the diffusively coupled Lorenz oscillators by comparing Figs.~\ref{fig:5}(a) and \ref{fig:5}(d), and is further supported by the parameter space shown in Fig.~\ref{fig:6}(d). Thus, a common effect of higher-order interactions in both oscillator models is to enable chimera states at weaker pairwise coupling.

\begin{figure}
    \centering
    \includegraphics[width=\linewidth]{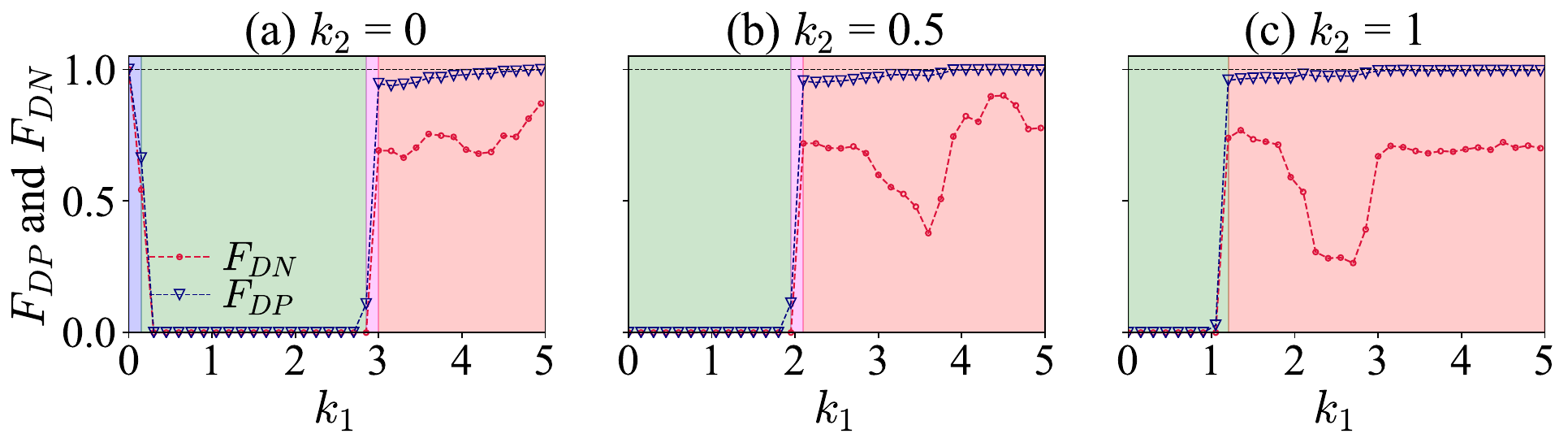}
    \includegraphics[width=\linewidth]{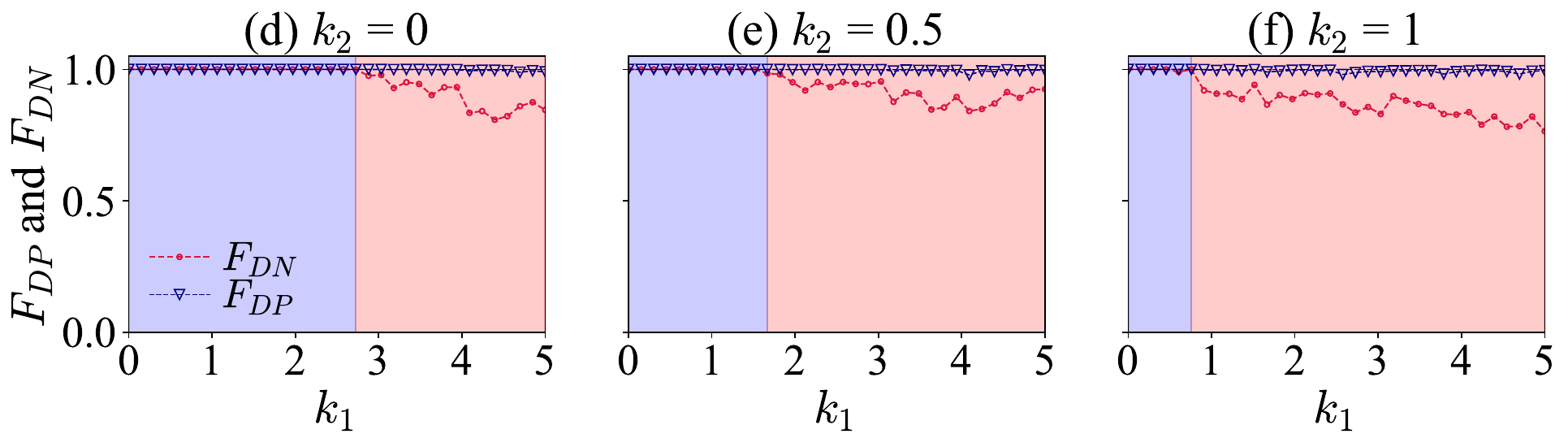}
  \caption{\textbf{ Statistical measures as functions of the pairwise coupling strength $k_1$.}} The average values of $F_{\mathrm{DP}}$ and $F_{\mathrm{DN}}$ are shown as functions of $k_1$ for coupled R\"{o}ssler oscillators in panels (a--c) and coupled Lorenz oscillators in panels (d--f) under diffusive coupling. The results correspond to $k_2=0$, $0.5$, and $1$ from left to right. Light green, blue, red, and magenta shaded regions denote synchronized, desynchronized, chimera, and synchronized cluster states, respectively, identified from the average values of $F_{\mathrm{DP}}$ and $F_{\mathrm{DN}}$. For both oscillator systems, increasing $k_2$ shifts the chimera region ($0<F_{\mathrm{DP}},F_{\mathrm{DN}}<1$) toward lower values of $k_1$, indicating that higher-order interactions facilitate the emergence of chimera states under diffusive coupling for this range.
    \label{fig:7}
\end{figure}

 This behavior is further supported by the average statistical measures $F_{\mathrm{DP}}$ and $F_{\mathrm{DN}}$, shown as functions of $k_1$ for increasing values of $k_2$ in Figs.~\ref{fig:7}(a--c) and \ref{fig:7}(d--f) for the R\"{o}ssler and Lorenz systems, respectively. As $k_2$ increases, the interval of $k_1$ satisfying $0<F_{\mathrm{DP}}<1$ and $0<F_{\mathrm{DN}}<1$ shifts toward smaller $k_1$, in agreement with the fractional basin stability results in Figs.~\ref{fig:4} and \ref{fig:5}. However, the influence of higher-order interactions is not universally monotonic and depends on the intrinsic oscillator dynamics. For the Lorenz system, larger values of $k_2$ continue to support chimera states over the parameter range explored, whereas for the R\"{o}ssler system, the desynchronized state becomes increasingly probable for $k_2 > 2$ when $k_1$ is also large. Therefore, although higher-order interactions consistently shift the onset of chimera states toward weaker pairwise coupling, increasing $k_2$ does not necessarily enhance their basin of attraction; the eventual collective state depends on the interplay between the intrinsic oscillator dynamics and the relative strengths of pairwise and higher-order interactions. We further note that, for some parameters within the chimera region, $F_{\mathrm{DP}}$ and $F_{\mathrm{DN}}$ approach unity but remain strictly below $1$, as indicated by the corresponding dashed lines in Fig.~\ref{fig:7}.

 Overall, the results demonstrate that higher-order interactions reorganize the dynamical-state landscape under diffusive coupling rather than universally promoting chimera states. Despite the structural and dynamical equivalence of the peripheral nodes imposed by the rotational symmetry of the wheel network, symmetry-broken states characterized by the coexistence of synchronized and desynchronized peripheral groups are sustained over different regions of the $(k_1,k_2)$ parameter space. The principal common effect of the 2-simplex interactions is a shift of the chimera region toward smaller $k_1$, whereas their influence at stronger $k_2$ is oscillator dependent, with chimera states remaining prevalent for the Lorenz system but competing increasingly with desynchronized states in the R\"{o}ssler system.

 \textbf{Conjugately coupled chaotic oscillators:}
 
By comparing Figs.~\ref{fig:4}(b) and \ref{fig:4}(e) for conjugately coupled R\"{o}ssler oscillators, we observe a pronounced effect of higher-order interactions on the emergence of chimera states. In the absence of higher-order interactions ($k_2=0$), chimera states are not observed, whereas their inclusion leads to the appearance of chimera states with high probability at relatively small values of $k_1$. With increasing $k_1$, however, their probability decreases and eventually vanishes beyond a critical value. For the conjugately coupled Lorenz oscillators (cf. Figs.~\ref{fig:5}(b) and \ref{fig:5}(e)), chimera states are already present at $k_2=0$ for sufficiently large $k_1$, while the inclusion of higher-order interactions modifies the range of $k_1$ over which they occur.

 The corresponding $(k_1,k_2)$ parameter spaces in Figs.~\ref{fig:6}(b) and \ref{fig:6}(e) further reveal that the influence of higher-order interactions is non-monotonic. Chimera states occupy substantial regions of the parameter space at intermediate values of $k_2$, whereas increasing $k_2$ further progressively favors synchronized dynamics. For the R\"{o}ssler system (cf. Fig.~\ref{fig:6}(b)), a broad region is also observed in which chimera and synchronized states have comparable probabilities, demonstrating a strong dependence of the asymptotic collective state on both coupling strengths and the initial conditions.

 Overall, these results show that higher-order interactions reorganize the dynamical-state landscape under conjugate coupling rather than universally promoting chimera states. At intermediate $k_2$, they extend the parameter region supporting symmetry-broken chimera states, whereas sufficiently strong higher-order coupling favors synchronization. The precise response is oscillator dependent: higher-order interactions are essential for the appearance of chimera states in the conjugately coupled R\"{o}ssler system within the explored parameter range, while in the Lorenz system they modify an already existing chimera regime. Thus, the emergence or suppression of chimera states results from the interplay between the intrinsic oscillator dynamics and the relative strengths of pairwise and higher-order interactions.

 \textbf{Mean-field diffusively coupled chaotic oscillators:}
 
For mean-field diffusively coupled R\"{o}ssler oscillators (cf. Figs.~\ref{fig:4}(c), \ref{fig:4}(f), and \ref{fig:6}(c)), chimera states are the most probable dynamical state only over a narrow range of $k_1$ when higher-order interactions are absent ($k_2=0$). With the inclusion of higher-order interactions, chimera states remain restricted to a relatively small region of the $(k_1,k_2)$ parameter space and are considerably less prevalent than under the diffusive and conjugate coupling schemes. Thus, for the R\"{o}ssler  system, mean-field diffusive coupling provides comparatively limited parameter regions supporting symmetry-broken chimera states.

 A markedly different behavior is observed for the Lorenz system (cf. Figs.~\ref{fig:5}(c), \ref{fig:5}(f), and \ref{fig:6}(f)), where chimera states occupy a substantially broader region of the $(k_1,k_2)$ parameter space. Their occurrence generally requires relatively stronger pairwise coupling, with chimera states becoming increasingly probable beyond a threshold value of $k_1$ for an appropriate range of $k_2$. In addition, synchronized cluster states also occupy a region of the parameter space and are particularly prominent for the Lorenz oscillators under mean-field diffusive coupling.
 
 Overall, these results demonstrate a strong dependence of the collective dynamics on the intrinsic oscillator model under mean-field diffusive coupling. Although the peripheral nodes remain structurally and dynamically equivalent, the local mean-field interactions modify the effective coupling experienced by each node and thereby alter the parameter regions supporting symmetry-broken states. The substantially different prevalence of chimera states in the R\"{o}ssler and Lorenz networks indicates that their occurrence cannot be attributed to the coupling architecture alone, but instead results from the interplay between the intrinsic oscillator dynamics and the relative strengths of the pairwise and higher-order mean-field interactions.

 \subsection{ Origin of Partial synchronization and Master Stability Function} \label{subsec:4.3}

 The wheel network possesses a highly symmetric topology in which the central hub is fixed and the peripheral nodes form a regular cycle. Consequently, the governing equations are equivariant under the dihedral symmetry group $D_{N-1}$, consisting of all rotations and reflections of the peripheral ring that leave the hub unchanged. A fundamental result from equivariant dynamical systems theory states that
symmetry subgroups generate invariant subspaces of the nonlinear dynamics
\cite{golubitsky2002symmetry,golubitsky2006nonlinear}. In particular, the reflection subgroup produces an exact partially synchronized invariant manifold in which pairs of peripheral nodes related by reflection remain synchronized, while the remaining degrees of freedom evolve independently. The existence of these invariant manifolds is therefore a purely structural consequence of the wheel topology and is independent of the oscillator model or coupling strengths. These manifolds provide the natural geometric framework within which partially synchronized solutions may evolve. (refer Supplementary Section S3.A for more details)

To support our numerical observations, we develop an analytical framework mapping the structural topology of the wheel network to the Master Stability Function (MSF)\cite{PecoraCarroll1998,Acharyya2026Comprehensive}. We define the local degree-normalized generalized graph Laplacians for the pairwise ($L^{(1)}$) and triadic ($L^{(2)}$) interactions as \cite{Gambuzza2021}:

\begin{equation}
\begin{aligned}
L^{(1)}_{ij} &= \delta_{ij} - \frac{A_{ij}}{d_i^1}, \\
L^{(2)}_{ii} &= 2, \\
L^{(2)}_{ij} &= -\frac{B_{ij}}{d_i^2}, \quad j \neq i.
\end{aligned}
\end{equation}

 where $B_{ij} = \sum_l A_{ijl}$ represents the number of triangles shared by nodes $i$ and $j$, while $d_i^1$ and $d_i^2$ denote the 1-simplex and 2-simplex degrees, respectively. Using these operators, the network's effective coupling matrix reduces to $M(k_1, k_2) = k_1 L^{(1)} + k_2 L^{(2)}$ (see Supplementary Section S3.B).

 When the variational equations are projected onto the transverse subspace, where the central hub remains synchronized and perturbations sum to zero across the peripheral nodes \cite{PecoraCarroll1998,pecora2014cluster}, the reduced and grounded operators exhibit an exact affine relationship (Supplementary Section S3.D):

\begin{equation}
\mathcal{L}^{(2)} = \mathcal{L}^{(1)} + I
\label{affine_relation}
\end{equation}

This exact topological identity proves that, on a wheel network, higher-order interactions contribute an independent, uniform additive spectral shift ($+I$). This fundamentally distinguishes the wheel topology from a standard globally coupled network, where triadic coupling merely acts as a redundant, proportional scalar to pairwise diffusion. The affine relation between the grounded pairwise and higher-order Laplacians (Eq.~\eqref{affine_relation}) plays a central role in the spectral analysis of the wheel network. Since the identity matrix shifts only the eigenvalues without changing the eigenvectors, both grounded Laplacians share the same discrete Fourier eigenvectors \cite{davis1994circulant,gray2006toeplitz}.As a result, the eigenvalues of the effective transverse coupling operator can be obtained analytically in closed form without numerical diagonalization (Supplementary Section~S2.D). The resulting transverse dispersion relation is

\begin{equation}
\Lambda_k
=
(k_1+2k_2)
-
\left(
\frac{2k_1}{3}+k_2
\right)
\cos\left(\frac{2\pi k}{N-1}\right)
\label{dispersion_relation}
\end{equation}

where $k=1,\ldots,N-2$ labels the discrete spatial modes (wavenumbers) of the peripheral ring.

Once the complete spectrum is known, the synchronization boundaries follow directly from the MSF. The synchronized state is linearly stable only if every non-zero eigenvalue lies inside the stable region of the MSF, where the largest transverse Lyapunov exponent is negative \cite{PecoraCarroll1998}. Consequently, the analytical expressions for $\Lambda_k$ and $\mu_H$ determine the synchronization boundaries explicitly in the $(k_1,k_2)$ parameter space. For the modified R\"{o}ssler oscillator with diffusive coupling, the resulting analytical synchronization boundary in the $(k_1,k_2)$ parameter plane is shown in Supplementary Fig.~S2 in Supplementary Section S3.D.

 The dispersion relation \eqref{dispersion_relation} also provides insight into how synchronization is lost. Since $\Lambda_k$ depends monotonically on the cosine term, the transverse eigenvalues are naturally ordered according to their spatial wavenumber. As the coupling parameters vary, the transverse modes approach the MSF stability boundary in a well-defined sequence. The mode that first crosses the MSF boundary is the first linearly unstable mode and that determines how the dynamics will evolve and deviate from synchronization.  Thus, the analytical spectrum predicts which spatial mode loses stability first and provides a mechanistic explanation for the onset of spatial pattern formation. The subsequent nonlinear evolution of this instability determines whether the network evolves toward desynchronized, chimera,  synchronized cluster states, as demonstrated by the numerical simulations.\\

\section{CONCLUSIONS}
\label{sec:5}
In this work, we investigated the emergence of collective dynamical states in wheel networks of coupled chaotic oscillators, with particular emphasis on the influence of higher-order interactions on chimera states. We considered R\"{o}ssler and Lorenz oscillators with 1-simplex  and 2-simplex interactions under diffusive, conjugate, and mean-field diffusive coupling schemes. By sampling multiple random initial conditions, we quantified the prevalence of synchronized, desynchronized, chimera, and synchronized cluster states through probability plots and constructed $(k_1,k_2)$ parameter-space diagrams by identifying the most probable dynamical state at each combination of pairwise and higher-order coupling strengths.

 Our results demonstrate that higher-order interactions substantially reorganize the collective dynamical landscape, but their influence on chimera states depends on both the coupling mechanism and the intrinsic oscillator dynamics. Under diffusive coupling, the principal common effect of higher-order interactions is to shift the chimera region toward smaller pairwise coupling strengths, allowing chimera states to occur at weaker $k_1$. At stronger higher-order coupling, however, the response becomes oscillator dependent, for example, chimera states remain prevalent for the Lorenz system, whereas desynchronized states increasingly compete with chimera states in the R\"{o}ssler system. Under conjugate coupling, chimera states are predominantly supported at intermediate higher-order coupling strengths, while sufficiently strong $k_2$ increasingly favors synchronization. In particular, higher-order interactions enable chimera states in the conjugately coupled R\"{o}ssler system within the explored parameter range, whereas they modify an already existing chimera regime in the Lorenz system. Mean-field diffusive coupling exhibits an even stronger dependence on the intrinsic oscillator dynamics, for instance, chimera states occupy a comparatively restricted parameter region for R\"{o}ssler oscillators, whereas Lorenz oscillators exhibit a substantially broader chimera region.

To quantitatively characterize these collective states, we introduced two statistical measures, $F_{\mathrm{DP}}$ and $F_{\mathrm{DN}}$, based on the fractions of desynchronized pairs and desynchronized nodes, respectively. Their joint use provides a systematic distinction among synchronized, desynchronized, chimera, and synchronized cluster states within the classification adopted in this work. In particular, these measures allow the coherent and incoherent subsets to be identified without prescribing their spatial locations in the network, making them useful for characterizing symmetry-broken states in the wheel topology.

 The emergence of chimera states among the structurally and dynamically equivalent peripheral nodes represents a spontaneous breaking of their dynamical symmetry. To complement the numerical observations, we performed a master stability function analysis for the diffusively coupled R\"{o}ssler system to determine the transverse stability of the completely synchronized state. The resulting stability boundaries are consistent with the numerically obtained $(k_1,k_2)$ dynamical regimes and identify parameter regions in which complete synchronization loses stability, thereby providing a stability-based framework for understanding the emergence of nonsynchronous collective dynamics.

 Overall, our results show that higher-order interactions neither universally promote nor suppress chimera states; rather, they reshape the dynamical-state landscape through their interplay with the coupling mechanism, and intrinsic oscillator dynamics. Within the parameter ranges explored, chimera states are most prevalent under diffusive coupling, followed by conjugate coupling, and are comparatively less prevalent under mean-field diffusive coupling. These findings highlight the nontrivial role of higher-order interactions in controlling symmetry-broken collective dynamics and emphasize that their effect cannot be inferred from interaction order alone. A complete theoretical analysis of the mechanisms governing the emergence and robustness of chimera states in networks with higher-order interactions, as well as the conditions under which such interactions promote or suppress these states, remains an important direction for future research.

\section*{Data Availability Statement}
No empirical data were analyzed in this study. All figures were generated from numerical simulations of the dynamical models described in the manuscript. The simulation codes used to produce the results are available from the corresponding author upon reasonable request.\\

\begin{acknowledgments}
MB and CM acknowledge support from the Anusandhan National Research Foundation (ANRF), India (Grant No. EEQ/2023/001080). CM further acknowledges support from the ANRF India (Grant No. SRG/2023/001846) and the Department of Science and Technology (DST), India (INSPIRE Faculty Grant No. IFA19-PH248).
\end{acknowledgments}

%\newpage

% \bibliography{sample}

%\newpage

\bibliographystyle{apsrev4-2}
\bibliography{sample}

\end{document}